\documentclass[11pt]{article}
\usepackage[a4paper,margin=1in]{geometry}
\usepackage[T1]{fontenc}
\usepackage[utf8]{inputenc}
\usepackage{lmodern}
\usepackage{microtype}
\usepackage{amsmath,amssymb}
\usepackage{booktabs}
\usepackage{graphicx}
\usepackage{xcolor}
\usepackage{enumitem}
\usepackage{verbatim}
\usepackage{fvextra}
\usepackage[round,authoryear]{natbib}
\usepackage[hidelinks]{hyperref}
\hypersetup{
  pdftitle={Learning the Platform, Not the Learner: Sequencing Shortcuts in the Offline Evaluation of Educational Recommender Systems and LLM Rankers},
  pdfauthor={Youssef Jdidou},
  pdfsubject={Offline evaluation of educational recommender systems},
  pdfkeywords={educational recommender systems, offline evaluation, sequential recommendation, exposure bias, curriculum sequencing, large language models, knowledge tracing, reproducibility}
}
\usepackage{caption}
\usepackage{tikz}
\usepackage{tcolorbox}
\usetikzlibrary{positioning,calc,arrows.meta}
\newcommand{\numDnovAfifteen}{+0.125}

\newcommand{\numDnovJunyi}{+0.048}

\newcommand{\numDoffAseventeen}{+0.024}

\newcommand{\numDoffEdNet}{+0.011}
\newcommand{\numDoffEedi}{+0.045}
\newcommand{\numDoffJunyi}{+0.039}

\newcommand{\numDonAseventeen}{+0.001}

\newcommand{\numDonEdNet}{+0.010}
\newcommand{\numDonEedi}{-0.002}
\newcommand{\numDonJunyi}{+0.044}

\newcommand{\numPtwoAfifteenP}{0.374}
\newcommand{\numPtwoAfifteenS}{0.232}

\newcommand{\numPtwoAninenTest}{158}

\newcommand{\numPtwoEdNetrho}{0.65}

\newcommand{\numPtwoSratioMax}{0.88}
\newcommand{\numPtwoSratioMin}{0.12}
\newcommand{\numPtwonStructMC}{seven}
\newcommand{\numPtwonsList}{A09 and EdNet}
\newcommand{\numPtworhoMax}{1.00}
\newcommand{\numPtworhoMedian}{0.91}
\newcommand{\numPtworhoMin}{0.62}
\newcommand{\numPtwosigList}{A15, A17, Junyi, Eedi and MOOC}
\newcommand{\numPtwoworseList}{A12}

\newcommand{\numSoffAnine}{0.033}
\newcommand{\numSoffAseventeen}{0.544}
\newcommand{\numSoffAtwelve}{0.319}

\newcommand{\numSoffEedi}{0.254}
\newcommand{\numSoffJunyi}{0.130}
\newcommand{\numSonAnine}{0.132}
\newcommand{\numSonAseventeen}{0.968}
\newcommand{\numSonAtwelve}{0.257}

\newcommand{\numSonEedi}{0.581}
\newcommand{\numSonJunyi}{0.748}

\newcommand{\numaucDktMax}{0.82}
\newcommand{\numaucDktMin}{0.65}

\newcommand{\numaucRateMax}{0.77}
\newcommand{\numaucRateMin}{0.50}

\newcommand{\numcorrSigList}{A12, A15, A17, EdNet, Junyi, Eedi and MOOC}
\newcommand{\numcorrSigN}{seven}
\newcommand{\numdCorrAfifteen}{+0.245}

\newcommand{\numdCorrMOOC}{+0.202}
\newcommand{\numdeltaAfifteen}{+0.102}

\newcommand{\numdeltaAseventeen}{+0.015}

\newcommand{\numdeltaEdNet}{+0.007}

\newcommand{\numdeltaJunyi}{+0.047}
\newcommand{\numdeltaMOOC}{+0.088}

\newcommand{\numfwdMax}{75}
\newcommand{\numfwdMin}{51}

\newcommand{\numitemsEdNet}{8,472}

\newcommand{\numletterHREediRandQwenSmall}{0.045}

\newcommand{\numletterShareEediRandQwenSmall}{84}

\newcommand{\numllmAnineHardPOP}{0.010}
\newcommand{\numllmAnineHardQwenLarge}{0.110}
\newcommand{\numllmAnineHardQwenSmall}{0.095}

\newcommand{\numllmAnineRandCONT}{0.650}

\newcommand{\numllmAnineRandLEX}{0.594}

\newcommand{\numllmAnineRandQwenSmall}{0.600}

\newcommand{\numllmBestEdNetHard}{0.120}

\newcommand{\numllmEdNetHardQwenSmall}{0.055}

\newcommand{\numllmEediHardPOP}{0.055}

\newcommand{\numllmMaxOverLex}{0.056}

\newcommand{\nummchroneAseventeen}{0.650}

\newcommand{\nummchroneMOOC}{0.393}

\newcommand{\numnDKTbelowPOP}{six}
\newcommand{\numnDatasets}{eight}

\newcommand{\numnDkt}{seven}
\newcommand{\numnDktBetter}{seven}
\newcommand{\numnNotSig}{three}

\newcommand{\numnSASRecBest}{seven}
\newcommand{\numnSig}{five}
\newcommand{\numnStructMC}{seven}

\newcommand{\numnZPDbelowPOP}{seven}

\newcommand{\numnovN}{six}
\newcommand{\numnovPmax}{0.162}
\newcommand{\numnovPmaxWhere}{A15}
\newcommand{\numnovSmax}{0.053}

\newcommand{\numnovshareAnine}{72}
\newcommand{\numnovshareAseventeen}{15}
\newcommand{\numnovshareAtwelve}{43}
\newcommand{\numnovshareEdNet}{66}

\newcommand{\numnsList}{A09 and Eedi}

\newcommand{\numptransAseventeen}{0.49}

\newcommand{\numquantAgree}{80}
\newcommand{\numquantMaxDiff}{0.070}
\newcommand{\numquantMeanDiff}{0.022}
\newcommand{\numquantN}{8}

\newcommand{\numrhoCorrMax}{1.00}
\newcommand{\numrhoCorrMedian}{0.66}
\newcommand{\numrhoCorrMin}{0.33}

\newcommand{\numrhoMax}{1.00}
\newcommand{\numrhoMedian}{0.95}

\newcommand{\numrhoMetaEedi}{0.61}
\newcommand{\numrhoMetaJunyi}{0.55}
\newcommand{\numrhoMetaMOOC}{0.59}
\newcommand{\numrhoMetaMax}{1.00}
\newcommand{\numrhoMetaMedian}{0.45}
\newcommand{\numrhoMetaMin}{0.20}
\newcommand{\numrhoMin}{0.81}
\newcommand{\numsameAnine}{63}
\newcommand{\numsameAseventeen}{40}
\newcommand{\numsameAtwelve}{69}
\newcommand{\numsameEdNet}{77}
\newcommand{\numsameEedi}{93}
\newcommand{\numsameJunyi}{60}

\newcommand{\numsasCapList}{A12, A15 and EdNet}

\newcommand{\numscaleGainMax}{+0.06}
\newcommand{\numscaleGainMin}{-0.03}
\newcommand{\numscaleLLMmaxHardAnineEdNet}{0.09}
\newcommand{\numscaleMCleadMin}{0.14}
\newcommand{\numscaleN}{100}
\newcommand{\numscaleSevenBeatsLex}{on none}

\newcommand{\numseedDmaxExtraEedi}{+0.043}

\newcommand{\numseedDminExtraEedi}{+0.024}

\newcommand{\numseedN}{five}

\newcommand{\numseedSdStableMax}{0.005}

\newcommand{\numseedStableList}{A15, A17, Junyi and MOOC}
\newcommand{\numseenSmax}{0.841}
\newcommand{\numseenSmin}{0.103}
\newcommand{\numsigDeltaMax}{0.102}
\newcommand{\numsigDeltaMin}{0.007}
\newcommand{\numsigList}{A15, A17, EdNet, Junyi and MOOC}

\newcommand{\numsimMCNeedAHF}{0.37}

\newcommand{\numsimRankNeedPlainAAA}{3}

\newcommand{\numsimRankNeedPlainAHF}{7}

\newcommand{\numsimRhoAAA}{0.53}

\newcommand{\numsimRhoAHF}{0.99}

\newcommand{\numsimSASRecNeedAAA}{1.00}

\newcommand{\numsimSASRecNeedAHF}{0.40}
\newcommand{\numsimSASRecNeedAJA}{0.35}

\newcommand{\numsimTauCorrAFA}{0.85}
\newcommand{\numsimTauCorrAHF}{0.75}
\newcommand{\numsimTauCorrAJA}{0.35}
\newcommand{\numsimTauPlainAAA}{0.86}

\newcommand{\numsimTauPlainAFA}{0.42}
\newcommand{\numsimTauPlainAHF}{0.24}
\newcommand{\numsimTauPlainAJA}{0.16}
\newcommand{\numsrcEdNetappchosethesetsprint}{0.71}

\newcommand{\numsrcEdNetreviewmodes}{0.95}

\newcommand{\numsrcJunyiopenedfromatopic}{0.92}
\newcommand{\numsrcJunyiplatformsuggestionmap}{0.89}

\newcommand{\numsrcLiftJunyichosenbythelearner}{0.027}

\newcommand{\numsrcLiftJunyiplatformsuggestionmap}{0.077}

\newcommand{\numsrcSJunyichosenbythelearner}{0.264}
\newcommand{\numsrcSJunyiopenedfromatopic}{0.658}
\newcommand{\numsrcSJunyiplatformsuggestionmap}{0.649}

\newcommand{\numsuccAseventeen}{5.2}

\newcommand{\numvAfifteenMCLD}{0.447}

\newcommand{\numvAfifteenSASRec}{0.522}

\newcommand{\numvEdNetMC}{0.035}

\newcommand{\numvEdNetSKNN}{0.037}

\newcommand{\numvalChanged}{one}

\newcommand{\numvalRhoMax}{1.00}
\newcommand{\numvalRhoMedian}{0.95}
\newcommand{\numvalRhoMin}{0.81}
\newcommand{\numvalSigList}{A15, A17, EdNet, Junyi and MOOC}
\newcommand{\numvalWorseList}{A12}
\newcommand{\numworseList}{A12}

\newcommand{\figmain}{%
\begin{figure}[t]
\centering
\includegraphics[width=\linewidth]{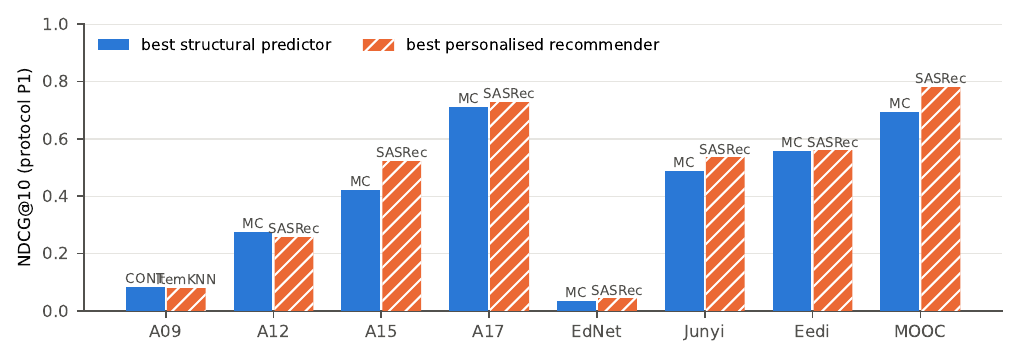}
\caption{Best structural predictor and best personalised recommender on each dataset (NDCG@10, protocol P1). The label above each bar names the method.}
\label{fig:main}
\end{figure}}

\newcommand{\figpaths}{%
\begin{figure}[t]
\centering
\includegraphics[width=\linewidth]{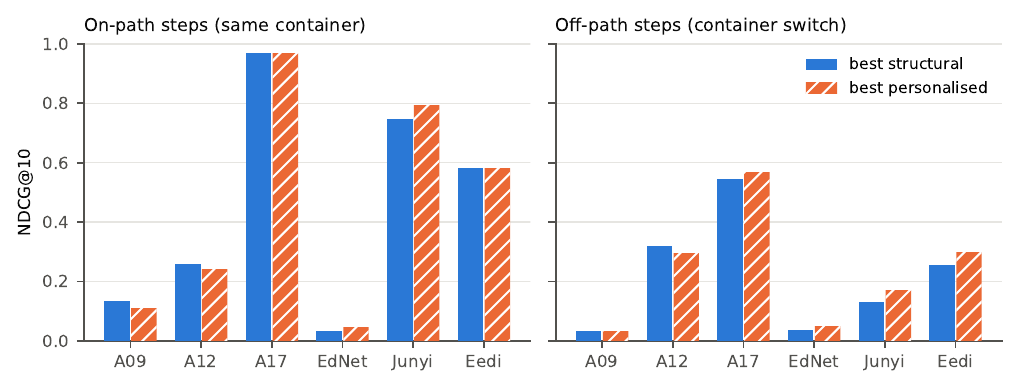}
\caption{NDCG@10 of the best structural predictor and of the best personalised recommender on on-path and off-path test steps (protocol P1). On four of the six datasets, accuracy drops sharply when the learner leaves the container of the last activity.}
\label{fig:paths}
\end{figure}}

\newcommand{\figllm}{%
\begin{figure}[t]
\centering
\includegraphics[width=\linewidth]{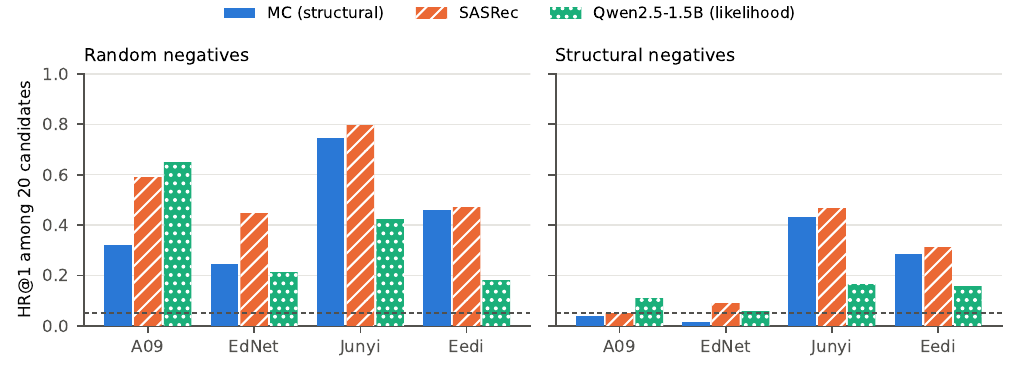}
\caption{HR@1 among 20 candidates for the transition heuristic MC, SASRec and the larger LLM (likelihood scoring), with random and with structural negatives. The dashed line is the expected value of a random ranking (0.05).}
\label{fig:llm}
\end{figure}}

\newcommand{\figteaser}{%
\begin{figure}[t]
\centering
\includegraphics[width=0.62\linewidth]{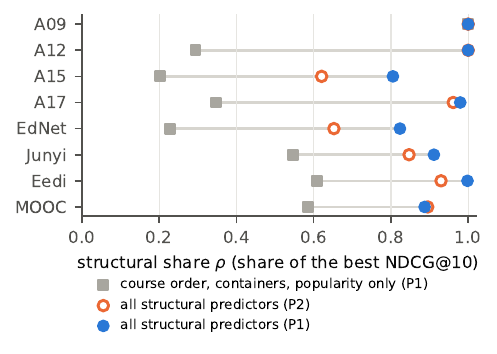}
\caption{Structural share $\rho$: the NDCG@10 of the best predictor that knows nothing about the learner except the last activity, as a share of the best NDCG@10 of any method, under leave-last-out (P1) and a global temporal split (P2). Grey squares: the same share without population transitions, from the course order, containers and popularity only.}
\label{fig:teaser}
\end{figure}}

\newcommand{\figsim}{%
\begin{figure}[t]
\centering
\includegraphics[width=0.92\linewidth]{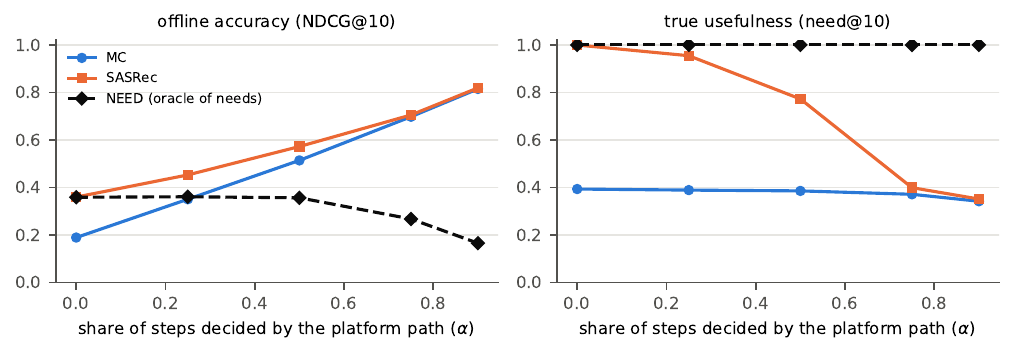}
\caption{Simulation with known needs: as the platform decides more steps, offline accuracy rewards imitation (left) while the true usefulness of the imitators stays low (right). NEED recommends exactly what each learner needs.}
\label{fig:sim}
\end{figure}}

\title{Learning the Platform, Not the Learner:\\[2pt]
\large Sequencing Shortcuts in the Offline Evaluation of Educational Recommender Systems and LLM Rankers}
\author{Youssef Jdidou\\[2pt]
\small École Nationale Supérieure d'Art et de Design (ENSAD)\\
\small Hassan II University of Casablanca, Morocco\\
\small \texttt{youssef.jdidou@gmail.com} \quad ORCID 0000-0002-7391-0940}
\date{}

\begin{document}
\maketitle

\begin{abstract}
Educational recommenders are usually compared offline, by how well they predict the next activity in logs of tutoring systems, learning applications and MOOCs, where the next activity is largely decided by the platform: teachers assign problem sets, tutors draw problems until mastery, engines select questions. On eight public datasets from five platforms, we compare four structural predictors, which use nothing about the learner beyond the last activity, with ten personalised recommenders and with open-weight LLM rankers from 0.5 to 7 billion parameters, under two protocols. The structural predictors reach between \numrhoMin{} and \numrhoMax{} of the best accuracy of any method (median \numrhoMedian{}), mostly through population transitions rather than the course order; the best personalised recommender is significantly better on \numnSig{} datasets, by \numsigDeltaMin{} to \numsigDeltaMax{} NDCG@10, and on Eedi in three of four training runs. Every method is far less accurate on steps along transitions never seen in training and on steps chosen by learners. LLM rankers look competent with random negatives but are never much better than a word-matching baseline, and with negatives that are plausible under the course structure they stay close to random or well behind the transition heuristic, whatever their size. In a simulation with known learner needs, the structural share tracks the share of steps decided by the platform, and offline accuracy ranks an oracle of the learners' needs below imitators of the platform. Weighting test steps by how unlikely the structure makes them restores much of the link with usefulness and, on the real data, reveals a significant personalisation signal on \numcorrSigN{} of the eight datasets. We propose a syllabus-aware evaluation checklist and release a Python package that reports the lift of any recommender over the structure.

\medskip\noindent\textbf{Keywords:} educational recommender systems; offline evaluation; sequential recommendation; exposure bias; curriculum sequencing; large language models; knowledge tracing; reproducibility
\end{abstract}

\section{Introduction}\label{sec:intro}
\figteaser

Recommender systems for learning propose the next exercise, problem set, video or quiz to a learner, and they are increasingly built with the same machinery as recommenders in other domains: collaborative filtering, sequential neural models and, most recently, large language models \citep{drachsler2015,cheng2026unier,ma2025llmcourse}. Most of them are developed and compared offline, on logs collected by tutoring systems, learning applications and MOOCs: a model sees what a learner has done so far and is rewarded when it ranks high the activity that the learner did next \citep{erdt2015}.

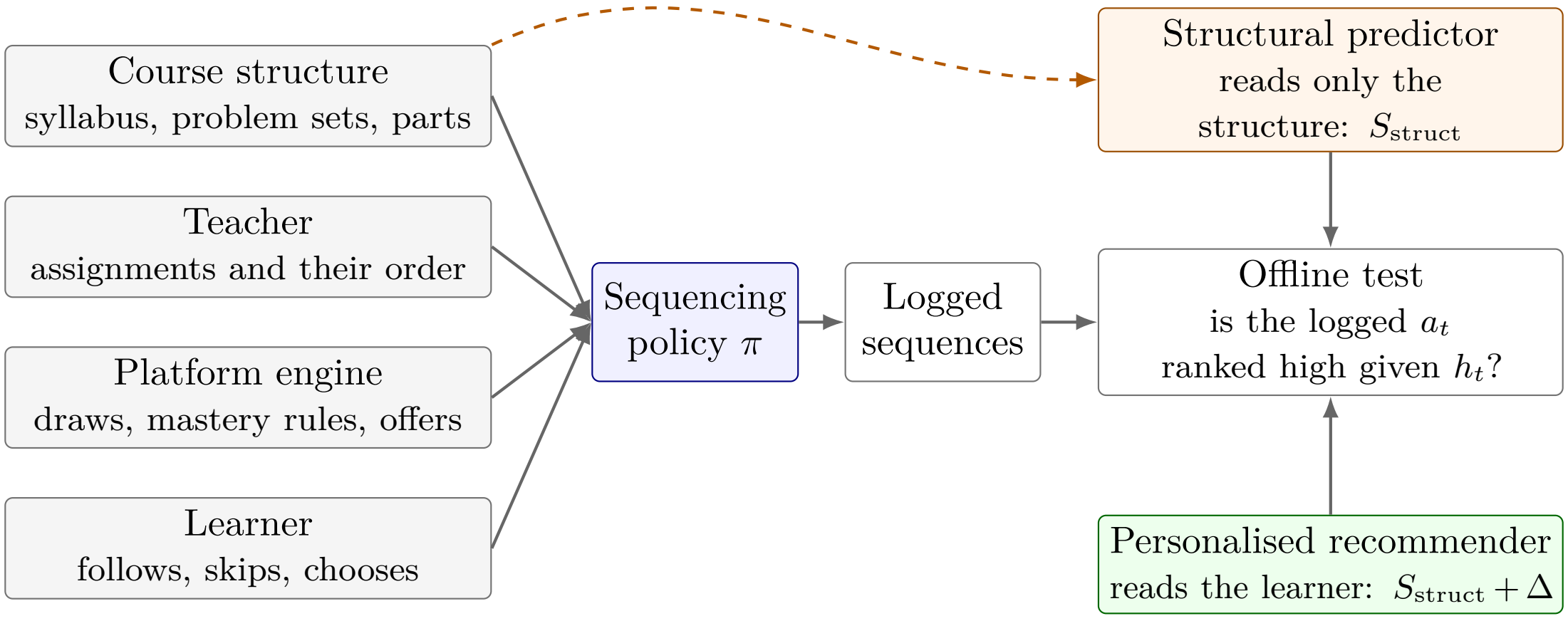
\begin{figure}[t]
\centering
\begin{tikzpicture}[
  font=\small,
  box/.style={draw=black!55, rounded corners=2pt, align=center, inner sep=3pt, fill=white},
  src/.style={box, text width=4.3cm, fill=black!4},
  arr/.style={-{Latex[length=2mm]}, thick, black!60}]
\node[src] (syl) at (0, 2.1) {Course structure\\\footnotesize syllabus, problem sets, parts};
\node[src] (tea) at (0, 0.7) {Teacher\\\footnotesize assignments and their order};
\node[src] (eng) at (0,-0.7) {Platform engine\\\footnotesize draws, mastery rules, offers};
\node[src] (lea) at (0,-2.1) {Learner\\\footnotesize follows, skips, chooses};
\node[box, fill=blue!6, draw=blue!50!black, minimum height=1.1cm, text width=1.7cm] (pi) at (4.15,0) {Sequencing\\policy $\pi$};
\node[box, minimum height=1.1cm, text width=1.6cm] (log) at (6.45,0) {Logged\\sequences};
\node[box, minimum height=1.1cm, text width=4.1cm] (test) at (10.05,0) {Offline test\\\footnotesize is the logged $a_t$ ranked high given $h_t$?};
\foreach \s in {syl,tea,eng,lea} \draw[arr] (\s.east) -- (pi.west);
\draw[arr] (pi) -- (log);
\draw[arr] (log) -- (test);
\node[box, text width=4.1cm, fill=orange!8, draw=orange!60!black] (g) at (10.05, 2.25) {Structural predictor\\\footnotesize reads only the structure: $S_{\text{struct}}$};
\node[box, text width=4.1cm, fill=green!7, draw=green!40!black] (f) at (10.05,-2.25) {Personalised recommender\\\footnotesize reads the learner: $S_{\text{struct}}+\Delta$};
\draw[arr] (g) -- (test);
\draw[arr] (f) -- (test);
\draw[arr, dashed, orange!70!black] (syl.north east) to[out=25, in=180] (g.west);
\end{tikzpicture}
\caption{Why offline next-activity accuracy partly measures the platform. The logged next activity is produced by a sequencing policy that combines the course structure, the teacher, the platform's engine and the learner. A predictor that reads only the structure already agrees with the logs to a degree $S_{\text{struct}}$; only the lift $\Delta$ beyond it can reflect personalisation, and even $\Delta$ measures agreement with $\pi$, not pedagogical value.}
\label{fig:concept}
\end{figure}

In educational logs, however, the next activity is seldom a free choice of the learner. A teacher assigns problem sets in a chosen order; a tutor draws the next problem from the current set until a mastery criterion is met; an application's engine selects the next question; a course has a syllabus. The logged sequence is the output of a \emph{sequencing policy} that the platform, the teacher and the curriculum mostly control (Figure~\ref{fig:concept}). A model can therefore score well offline by reproducing that policy, without learning anything about what a particular learner needs. This is a shortcut in the sense of \citet{geirhos2020shortcut}: a decision rule that succeeds on the benchmark but not on the intended task. Recent work found such shortcuts in general sequential-recommendation benchmarks, where a training-free transition heuristic matches modern generative recommenders \citep{han2026shortcut}. In educational data the shortcut is not an accident of dataset construction; it is the design of the platform. Early MOOC studies already noted that following the syllabus predicts a sizeable share of next resources \citep{tang2016mooc}, but no study has measured how much of the reported accuracy of educational recommenders the structure explains, on which steps personalisation adds anything, and what this implies for LLM-based recommenders, which are evaluated on small candidate sets where the choice of negatives is decisive \citep{krichene2020,hou2024zeroshot}.

This paper asks a simple question: when a recommender or an LLM scores well on such logs, has it learned the learner or the platform? Put differently, \emph{can recommenders and LLMs beat the syllabus?}, where by the syllabus we mean, broadly, the sequencing that the course and the platform impose: the course order, the grouping of activities into problem sets, parts or topics, and the typical paths that assignments produce. We audit offline next-activity evaluation on eight public datasets from five platforms (ASSISTments, the Santa TOEIC application of EdNet, Junyi Academy, Eedi and a MOOC), covering teacher-assigned problem sets, an adaptive engine, a self-paced knowledge map, teacher-assigned quizzes and a course. We compare four structural predictors, which read nothing about the learner beyond the last activity, with ten personalised recommenders (collaborative filtering, session-based, peer-based, knowledge tracing and sequential) and with open-weight LLM rankers, under two evaluation protocols and two kinds of candidate negatives. We summarise each benchmark by the accuracy available from the structure alone, $S_{\text{struct}}$, the lift of each recommender over it, $\Delta$, and the structural share $\rho$ (Section~\ref{sec:framing}).

Our findings are the following.
\begin{itemize}[leftmargin=*,itemsep=2pt]
\item \textbf{The structure explains most of the measured accuracy} (Figure~\ref{fig:teaser}). The structural share ranges from \numrhoMin{} to \numrhoMax{} (median \numrhoMedian{}). The best personalised recommender is significantly better than the best structural predictor on \numnSig{} of the \numnDatasets{} datasets, by \numsigDeltaMin{} to \numsigDeltaMax{} NDCG@10, and not on the other \numnNotSig{} (on Eedi, it is in three of four training runs).
\item \textbf{The shortcut is the paths, not the syllabus as written.} The best structural predictor is almost always the population transition heuristic. Without it, the course order, containers and popularity reach a median of only \numrhoMetaMedian{} of the best accuracy.
\item \textbf{Novel, off-path and learner-chosen steps are where the task is hard.} On steps along transitions never seen in training, no method reaches an NDCG@10 above \numnovPmax{} on the \numnovN{} datasets with enough such steps; on four of the six datasets with containers, the best structural and personalised methods are far less accurate on steps that leave the current container; and on Junyi they predict exercises chosen by learners less than half as well as exercises suggested by the platform. Personalised methods do not close these gaps, and pedagogical policies based on knowledge tracing rank below popularity even with a deep model of the learner.
\item \textbf{LLM rankers look competent only where the task is easy.} On candidate sets with random negatives, an open LLM can look strong, but a baseline that only matches words with the last activity's name does as well. With negatives that the course structure finds plausible, every method loses accuracy: the LLMs fall close to the random level on two datasets and stay well behind the transition heuristic on the other two. Going from 0.5 to 7 billion parameters changes none of this.
\item \textbf{Offline accuracy can rank the right recommender last, and a structure-corrected view helps.} In a simulation with known needs, the structural share tracks the share of steps the platform decides, and once it decides most steps, NDCG@10 ranks an oracle of the learners' needs below the imitators of the platform. Weighting test steps by the inverse of their structural probability restores much of the link with usefulness; on the real data, it shows that personalised recommenders do beat the structure on the steps the structure does not explain, on \numcorrSigN{} of the \numnDatasets{} datasets.
\end{itemize}

The contributions are (i) a framing of offline next-activity evaluation as imitation of a sequencing policy, with three summary quantities that separate structure from personalisation; (ii) a cross-platform audit of fourteen recommenders and open-weight LLMs on eight datasets, with protocols that avoid the usual pitfalls (exact handling of ties, new-activity targets, global temporal splits); (iii) a simulation with known needs that validates these quantities, and a structure-corrected accuracy (Section~\ref{sec:sim}); (iv) a syllabus-aware evaluation checklist (Section~\ref{sec:protocolrec}); and (v) an installable Python package that reports the lift of any recommender over the structure, with all the code and per-case results of this study \citep{jdidou2026code}.

\begin{tcolorbox}[colback=black!3,colframe=black!45,boxrule=0.4pt,arc=1pt,left=6pt,right=6pt,top=4pt,bottom=4pt,title={\small\textbf{Takeaways for anyone reporting offline results of an educational recommender}},coltitle=black,colbacktitle=black!8,fonttitle=\small]
\small
\begin{enumerate}[leftmargin=*,itemsep=1pt,topsep=1pt]
\item Score popularity, the course order, container continuation and the transition heuristic on exactly the same cases. A score means little until it is reported as a lift over the best of them.
\item Break results down by kind of step: seen and novel transitions, steps inside and outside the current container, and, if the log records it, steps started by the platform or by the learner. Personalisation matters on the hard ones.
\item Do not evaluate LLM rankers on random negatives alone: add structurally plausible negatives and a word-matching baseline.
\item Repeat the evaluation with a global temporal split.
\end{enumerate}
\end{tcolorbox}

\section{Related work}\label{sec:related}

\paragraph{Baselines and pitfalls in offline evaluation.}
A series of reproducibility studies has shown that progress claimed for new recommendation algorithms often disappears against well-tuned simple baselines. \citet{ferraridacrema2019,ferraridacrema2021} could reproduce only a minority of recent neural approaches, and most of those were outperformed by nearest-neighbour or linear methods; \citet{rendle2020ncf} found that a carefully tuned dot product beat a learned similarity; and \citet{ludewig2021} reported that nearest-neighbour heuristics were in most cases preferable to neural session-based recommenders. Other work has examined the evaluation procedure itself: sampled metrics can rank methods inconsistently with full rankings \citep{krichene2020}, splitting data without respecting the global timeline leaks future information \citep{ji2023leakage}, and many offline studies use flawed datasets, splits or metrics \citep{hidasi2023flaws}. Recorded interactions are themselves shaped by the system that produced them: recommendations act as treatments whose exposure biases the logs \citep{schnabel2016}, feedback loops make the data more homogeneous \citep{chaney2018}, and offline protocols built on such logs need to be interpreted with care \citep{jeunen2019}. Our study applies this line of scrutiny to a domain in which the exposure is not incidental but designed.

\paragraph{Shortcut-solvable benchmarks.}
Shortcut learning describes models that succeed on a benchmark through decision rules that do not reflect the intended task and that fail to transfer \citep{geirhos2020shortcut}. In recommendation, \citet{han2026shortcut} recently showed that a training-free transition-graph heuristic, using only the last one or two items, matches or beats modern generative recommenders on many standard sequential benchmarks, and traced this to low-branching transitions, feature-smooth transitions and weak dependence on long histories. Educational logs are an extreme case of this situation, because their transitions are produced on purpose by curricula, teachers and adaptive engines. To our knowledge, no study has quantified this source of predictability across educational datasets, separated it from personalisation, or examined how it affects LLM-based recommenders.

\paragraph{Evaluating educational recommenders.}
Recommender systems for technology-enhanced learning have been reviewed repeatedly \citep{manouselis2011,drachsler2015}, and surveys of their evaluation document a wide variety of methods and goals beyond accuracy \citep{erdt2015}. Instructional sequencing has also been studied as a policy-learning problem: the robust evaluation matrix compares instructional policies under several student simulators \citep{doroudi2017rem}, and a review of reinforcement learning for instructional sequencing examines when data-driven sequencing policies have improved on simpler baselines \citep{doroudi2019reward}. On MOOC clickstreams, \citet{tang2016mooc} observed that simply following the syllabus already predicted the next resource with an accuracy of 23\% on average, and \citet{pardos2017} deployed a next-step recommendation framework in a live MOOC. Exercise recommendation is now benchmarked at scale: UniER compares 18 item-level and path-level methods on nine datasets with a simulated cognitive-gain metric \citep{cheng2026unier}, and recent methods combine knowledge tracing \citep{corbett1994} with semantic information and reinforcement learning \citep{ozyurt2025exrec}. Peer-based methods such as MCLD recommend the activities passed by the classmates whose record is closest to the learner's \citep{jdidou2016,jdidou2021}. These works evaluate methods; we evaluate the evaluation, by asking how much of the measured accuracy any method could obtain from the platform's sequencing alone.

\paragraph{LLMs as recommenders and in education.}
Large language models can rank candidate items zero-shot, but they are sensitive to the order of the history and biased by popularity and by the position of candidates in the prompt \citep{hou2024zeroshot}; more generally, they prefer certain option labels in multiple-choice formats \citep{zheng2023mcq}. In education, LLMs have been evaluated as course recommenders on MOOC data, with accuracy comparable to traditional models \citep{ma2025llmcourse}, while comparisons with knowledge tracing caution that LLMs alone assess learners' evolving knowledge less reliably \citep{hooshyar2025llm}. LLM rankers are almost always evaluated on small candidate sets, which makes the choice of negative candidates decisive \citep{krichene2020}; we show that in educational data this choice determines whether the course structure alone can solve the task.

\section{Next-activity prediction as imitation of a sequencing policy}\label{sec:framing}

\paragraph{Setting.}
A learner's log is a sequence of steps $(a_1,o_1),(a_2,o_2),\dots$, where $a_t$ is an activity (a problem, an exercise, a set of questions, a quiz) and $o_t\in\{0,1\}$ indicates whether it was passed. The history before step $t$ is $h_t=((a_1,o_1),\dots,(a_{t-1},o_{t-1}))$. A recommender $f$ maps a history to a ranking of the activities. In the standard offline protocol, $f$ is rewarded when the activity $a_t$ that the learner actually did ranks high in $f(h_t)$, according to a metric $M$ such as the hit rate or the normalised discounted cumulative gain at a cut-off $K$.

\paragraph{Who chose $a_t$?}
In educational platforms, $a_t$ is rarely chosen by the learner alone. It is the output of a \emph{sequencing policy} $\pi$ that combines several decision makers: the course designer who fixes a syllabus, the teacher who assigns problem sets in a chosen order, the platform that draws the next problem from a pool or selects it adaptively, and the learner, who may follow, skip or revisit. Offline next-activity evaluation therefore measures how well $f$ \emph{agrees with} $\pi$ on logged histories. This is not the quantity of interest (whether the activities recommended by $f$ would help learners), and it can be large for reasons unrelated to the learner. Whenever $\pi$ depends on the course structure, a predictor that reads only that structure can agree with $\pi$ without knowing anything about the learner.

\paragraph{Structural predictors and lift over structure.}
We call \emph{structural} a predictor whose ranking depends on the history only through the last activity and statistics shared by all learners: popularity, the course order, membership of the same container (problem set, skill, part, topic, scheme of work) and population transition frequencies (Section~\ref{sec:methods}). Let $\mathcal{G}$ be a set of structural predictors and $\mathcal{F}$ a set of personalised recommenders. We report three quantities for each dataset:
\begin{align}
S_{\text{struct}} &= \max_{g\in\mathcal{G}} M(g), &
\Delta(f) &= M(f)-S_{\text{struct}}, &
\rho &= \frac{S_{\text{struct}}}{\max_{h\in\mathcal{G}\cup\mathcal{F}} M(h)}.
\end{align}
$S_{\text{struct}}$ is the accuracy available without personalisation, $\Delta(f)$ is the \emph{lift over structure} of a recommender, and $\rho$, the \emph{structural share}, is the fraction of the best attainable offline accuracy that the course structure alone delivers. A benchmark with $\rho$ close to one mostly rewards replaying the platform; the evidence it provides about personalisation lies in $\Delta$, not in $M$. Note that $\Delta$ is itself measured by agreement with $\pi$: a positive lift shows that a model predicts the logged choices better than the structure does, not that its recommendations are pedagogically better.

\paragraph{On-path and off-path steps.}
To locate where personalisation could matter, we split test steps by whether the next activity stays in the container of the last one (\emph{on-path}) or leaves it (\emph{off-path}). On-path steps are typically decided by the platform or the assignment (the next problem of a problem set, the next question of a part); off-path steps are those where a teacher, the platform or the learner switches to something else, which is exactly where a recommender would have to make a choice. When the logs record who initiated an activity, as EdNet and Junyi do, we also compare activities started by the platform with activities chosen by the learner.

\section{Experimental setup}\label{sec:setup}

\subsection{Datasets}\label{sec:data}
We use eight public datasets from five platforms (Table~\ref{tab:datasets}). They cover the main ways in which the next activity is decided: teacher-assigned problem sets with problems drawn by the tutor, a mobile application whose engine selects the questions, a self-paced platform organised as a knowledge map, quizzes assigned by teachers within schemes of work, and the course of a MOOC.
\begin{description}[leftmargin=0pt,itemsep=2pt,font=\normalfont\bfseries]
\item[A09.] ASSISTments 2009--2010 skill builders, original release \citep{feng2009,heffernan2014}. Activities are problems; the container is the problem set, from which the tutor draws problems until a mastery criterion is met. We keep main problems, remove the rows duplicated by multi-skill tagging, and order steps by the global response order.
\item[A12, A15, A17.] ASSISTments 2012--2013, 2015 and the 2017 longitudinal data, in the versions prepared by \citet{gervet2020}. In A12 and A17 activities are problems and, since these versions do not include problem-set identifiers, the skill serves as container (ASSISTments skill builders are organised by skill). In A15 activities are the 100 skill-builder problem sets themselves and no container is defined.
\item[EdNet.] Logs of Santa, a mobile application preparing learners for the TOEIC English test \citep{choi2020ednet}, in the most detailed format (KT4), for a public sample of 4,396 learners (4,387 after preprocessing). Activities are \emph{bundles}, the sets of one to five questions that the application presents together; the container is the TOEIC part (1 to 7). Each bundle records how it was started: in \emph{sprint} mode the learner chooses a part and the application serves its questions; \emph{adaptive offers} and \emph{tutor} recommendations are initiated by the application; review modes revisit past questions.
\item[Junyi.] Junyi Academy 2015, a mathematics platform built on Khan Academy's open-source code and organised as a knowledge map \citep{chang2015junyi}; we draw a random sample of 20,000 learners. Activities are exercises; the container is the topic of the map. Each attempt records whether the exercise was \emph{suggested} by the platform from the prerequisite map, opened from a \emph{topic} icon, done in \emph{review} mode after proficiency, or chosen by the learner (\emph{self}).
\item[Eedi.] Diagnostic questions from the NeurIPS 2020 Education Challenge (tasks 3 and 4) \citep{wang2020eedi}. Answers are time-stamped to the minute, which leaves the order of questions within a quiz undetermined, so activities are quizzes assigned by teachers; the container is the scheme of work in which the quiz was assigned.
\item[MOOC.] The ACT-MOOC data \citep{kumar2019jodie}, derived from the KDD Cup 2015 dropout-prediction data: the actions of 7,047 learners on the 97 activities of one MOOC, time-stamped to the second (ties are broken by the action identifier). Activities are anonymised and the log records neither outcomes nor course structure, so every step counts as passed, CONT reduces to ORDER, and the two methods that need outcomes (ZPD and DKT, below) are not applied.
\end{description}
In every dataset, consecutive responses of a learner to the same activity are merged into one \emph{step}, which is passed when at least half of the responses are correct. Activities with fewer than five steps are removed. A public preprocessed version of OLI Engineering Statics was examined and discarded: its items appear to be problem steps, as 74\% of transitions go to the next item identifier, so the next item is mostly fixed by the problem itself rather than chosen.

\subsection{Task, protocols and metrics}\label{sec:protocol}
\paragraph{Next new activity.} Given a history, a method ranks all activities the learner has not done yet, and the target is the next activity the learner started that was new to them. Activities already in the history are removed from the ranking of every method. This matches the purpose of an activity recommender (proposing what to do next) and removes the trivial gains of re-recommending past activities.

\paragraph{Protocol P1 (leave-last-out).} For each learner, the last new step is the test target and the previous new step the validation target; models are trained on all learners' sequences cut just before their validation target. This is the protocol of most sequential recommendation studies, and the one under audit. Hyperparameters are selected on (at most 4,000) validation cases by NDCG@10, and the selected configuration is evaluated on all test cases.

\paragraph{Protocol P2 (global temporal split).} Because leave-last-out mixes time periods across learners \citep{ji2023leakage}, we repeat the study with a global cut-off at the 80th percentile of step times: models are trained on all steps before the cut-off, and each learner active on both sides contributes one case whose target is their first new activity after it. Hyperparameters are selected with an earlier cut-off (70th percentile) on the training period only.

\paragraph{Metrics.} We report HR@$K$ (whether the target is among the first $K$ activities) and NDCG@10 over the full ranking of new activities. Several structural predictors give equal scores to many activities, so every metric is computed as its exact expectation under a uniformly random order of tied activities; ties are never broken by activity index. Intervals are 95\% percentile bootstrap intervals over test cases (2,000 resamples); differences between two methods are tested with a paired bootstrap and Holm's correction across datasets.

\subsection{Methods}\label{sec:methods}
\paragraph{Structural predictors.} \textbf{POP} ranks activities by the number of training learners who did them. \textbf{ORDER} follows the course order: each activity receives the median relative position of its first occurrence in the training learners' sequences, and the activities after the last one are ranked nearest first, followed by the earlier ones. \textbf{CONT} ranks the unseen activities of the last activity's container first (in course order), then the others as ORDER. \textbf{MC} ranks activities by how often they followed the last activity in the training sequences (a first-order Markov chain with a popularity back-off), which captures typical paths without any course metadata. MC is a classical sequential baseline; we count it as structural because it is not personalised: all learners whose last activity is the same receive the same ranking. None of the four uses anything about the learner beyond the last activity.

\paragraph{Personalised recommenders.} Collaborative filtering: item-based nearest neighbours with recency weighting (\textbf{ItemKNN}; \citealp{sarwar2001}), \textbf{RP3beta} \citep{paudel2016rp3}, \textbf{EASE} \citep{steck2019ease} and implicit alternating least squares with fold-in of the full history (\textbf{iALS}; \citealp{hu2008ials}). Session-based: \textbf{SKNN}, in which the $k$ training learners whose activities overlap most with the learner's $m$ most recent distinct activities (recent ones weighted more) vote for the activities they did, weighted by similarity; this is the vector variant of session-based nearest neighbours, which was competitive with or better than neural session-based models in the comparisons of \citet{jannach2017rnnknn} and \citet{ludewig2018evaluation}. Peer-based: \textbf{MCLD}, which recommends the activities passed by the most-coincident, least-different classmates \citep{jdidou2016,jdidou2021}, and its adaptive variant \textbf{aMCLD}, which takes the $k$ first peers in the same order; here all training learners act as classmates. Knowledge tracing: \textbf{ZPD} estimates the learner's success rate on each skill from the history and recommends the unseen activities whose predicted success is closest to a target rate $\tau$, in the spirit of keeping practice at a desirable difficulty \citep{wilson2019eighty}; it represents pedagogically motivated policies, which do not try to imitate the logs. \textbf{DKT} applies the same policy with the success probabilities of deep knowledge tracing \citep{piech2015dkt}, an LSTM over (skill, outcome) steps, so that a weak success estimate cannot explain the policy's results; on the P1 test cases, DKT predicts the outcome of the target step better than ZPD's smoothed rates on \numnDktBetter{} of the \numnDkt{} datasets with outcomes (AUC \numaucDktMin{}--\numaucDktMax{} against \numaucRateMin{}--\numaucRateMax{}). Sequential: \textbf{SASRec} \citep{kang2018sasrec} with outcome-aware input tokens (activity and pass/fail) and a cross-entropy loss \citep{klenitskiy2023}. All personalised methods are tuned on the validation cases; grids and implementation details are given in Appendix~\ref{app:hparams}. RAND (random ranking) is reported for reference.

\subsection{LLM rankers}\label{sec:llmsetup}
We evaluate open-weight instruction-tuned models of four sizes from one family, which share a tokenizer, so that every size reads exactly the same prompts: Qwen2.5-0.5B-Instruct and Qwen2.5-1.5B-Instruct \citep{qwen2024}, run in bfloat16 with Hugging Face Transformers \citep{wolf2020transformers}, and Qwen2.5-3B-Instruct and Qwen2.5-7B-Instruct, run 4-bit quantized (the official Q4\_K\_M files) with llama.cpp \citep{llamacpp}. All run on ordinary CPUs, so that the experiment can be reproduced without special hardware or paid services. Since an LLM cannot rank thousands of activities in one prompt, all methods are compared on candidate sets of 20 activities: the target and 19 negatives the learner has not done. We use two kinds of negatives. \emph{Random} negatives are drawn uniformly. \emph{Structural} negatives are drawn from the 20 activities ranked highest by MC and the 20 ranked highest by CONT, so that every candidate is plausible from the course structure alone. By construction they penalise MC and CONT, whose own top-ranked activities they are; their purpose is to test the other methods on decisions that the structure cannot make.

The LLM scores candidates by likelihood. The prompt describes the platform, lists the learner's ten most recent activities with their names and outcomes, and ends with ``The next activity the student starts is:''. The score of a candidate is the mean log-probability of its name as the continuation, minus the mean log-probability of the same name after a prompt that contains no history, a pointwise mutual information that removes the model's preference for short or frequent names. Activity names are those provided by the datasets: skill names for A09, part names for EdNet, exercise and topic names for Junyi and subject names for Eedi (the other ASSISTments versions provide only identifiers and are not used). We first tried the common multiple-choice format, with the candidates labelled A to T and the score of a candidate given by the probability of its letter, but the smallest model then chose the same letter in most cases whatever the content (\numletterShareEediRandQwenSmall{}\% of the Eedi cases), the label-selection bias documented by \citet{zheng2023mcq}; its accuracy was at the random level (HR@1 of \numletterHREediRandQwenSmall{}), and the likelihood formulation avoids the problem. Each dataset contributes 200 test cases per type of negatives, sampled at random from the P1 test cases, and every other method is scored on exactly the same candidates; the 3B and 7B models, which take up to a minute per case on a laptop CPU, were scored on the first \numscaleN{} of them. The llama.cpp scorer reproduces the bfloat16 scorer: with the 0.5B model in 16-bit precision, it puts the same candidate first in 40 of 40 test cases, and with the 0.5B model quantized like the larger ones, HR@1 changes by \numquantMeanDiff{} on average and at most \numquantMaxDiff{} over the \numquantN{} candidate sets (same top candidate in \numquantAgree{}\% of cases). Appendix~\ref{app:prompt} gives an example prompt.

\section{Results}\label{sec:results}

\begin{table}[t]
\centering
\caption{Datasets after preprocessing. Steps merge consecutive responses to the same activity. \emph{Median}: median number of steps per learner. \emph{Test cases}: learners with a test target under protocol P1. \emph{Revisits}: share of steps that return to an activity already done (these steps stay in the histories but are never targets).}
\label{tab:datasets}
\scriptsize
\setlength{\tabcolsep}{4pt}
\resizebox{\linewidth}{!}{\begin{tabular}{llllrrrrrr}
\toprule
Dataset & Platform & Activity & Container & Learners & Activities & Steps & Median & Test cases & Revisits \\
\midrule
A09 & ASSISTments & problem & problem set & 4,197 & 12,731 & 263,040 & 19 & 3,877 & 2\% \\
A12 & ASSISTments & problem & skill & 22,574 & 42,903 & 2,660,864 & 58 & 22,564 & 5\% \\
A15 & ASSISTments & problem set & none & 14,228 & 100 & 154,599 & 7 & 12,668 & 25\% \\
A17 & ASSISTments & problem & skill & 1,708 & 2,618 & 446,497 & 227 & 1,708 & 13\% \\
EdNet & Santa (EdNet) & question bundle & TOEIC part & 4,387 & 8,472 & 1,975,317 & 164 & 4,367 & 13\% \\
Junyi & Junyi Academy & exercise & map topic & 20,000 & 699 & 779,740 & 8 & 14,006 & 67\% \\
Eedi & Eedi & quiz & scheme of work & 4,918 & 92 & 123,938 & 21 & 4,915 & 19\% \\
MOOC & MOOC (KDD Cup 2015) & course activity & none & 7,047 & 97 & 320,984 & 30 & 7,008 & 46\% \\
\bottomrule
\end{tabular}
}
\end{table}

\subsection{RQ1: How much of the next activity does the structure predict?}\label{sec:rq1}
Table~\ref{tab:diagnostics} describes the test transitions. In five of the six datasets with containers, most next activities stay in the container of the last one, from \numsameJunyi{}\% of steps in Junyi to \numsameEedi{}\% in Eedi (\numsameAnine{}\% in A09, \numsameAtwelve{}\% in A12 and \numsameEdNet{}\% in EdNet); in A17, \numsameAseventeen{}\% do. Transitions are concentrated: in A17 the last activity has a median of \numsuccAseventeen{} effective successors and the logged next activity has a mean training transition probability of \numptransAseventeen{}, so the first-order transition heuristic MC alone places the exact next problem first in \nummchroneAseventeen{} of the test cases; in the MOOC it does so in \nummchroneMOOC{} of the cases. The course order is a much weaker signal than transitions: only between \numfwdMin{}\% and \numfwdMax{}\% of next activities lie ahead of the last one in the estimated course order, and ORDER is far below MC on every dataset (Table~\ref{tab:main}).

\begin{table}[t]
\centering
\caption{How predictable the next new activity is from the course structure (protocol P1, test steps). \emph{Same container}: next activity in the container of the last one. \emph{Forward}: next activity at or after the last one in the course order. \emph{Successors}: median effective number of successor activities of the last activity in the training data (exponential of the entropy of its transition distribution). $P(a_t\mid a_{t-1})$: mean training transition probability of the actual next activity. $S_{\text{struct}}$: NDCG@10 of the best structural predictor.}
\label{tab:diagnostics}
\small
\resizebox{\linewidth}{!}{\begin{tabular}{lrrrrrrl}
\toprule
Dataset & Same container & Forward & Successors & $P(a_t\mid a_{t-1})$ & MC HR@1 & $S_{\text{struct}}$ & Best structural \\
\midrule
A09 & 63\% & 52\% & 22.0 & 0.03 & 0.038 & 0.083 & CONT \\
A12 & 69\% & 54\% & 37.6 & 0.15 & 0.214 & 0.276 & MC \\
A15 & -- & 56\% & 26.0 & 0.08 & 0.249 & 0.420 & MC \\
A17 & 40\% & 60\% & 5.2 & 0.49 & 0.650 & 0.713 & MC \\
EdNet & 77\% & 51\% & 222.7 & 0.01 & 0.013 & 0.035 & MC \\
Junyi & 60\% & 64\% & 19.0 & 0.15 & 0.367 & 0.488 & MC \\
Eedi & 93\% & 67\% & 13.3 & 0.12 & 0.323 & 0.558 & MC \\
MOOC & -- & 75\% & 8.7 & 0.18 & 0.393 & 0.692 & MC \\
\bottomrule
\end{tabular}
}
\end{table}

\subsection{RQ2: Do personalised recommenders beat the structure?}\label{sec:rq2}
\begin{table}[t]
\centering
\caption{NDCG@10 for the next new activity under protocol P1 (full ranking of all activities the learner has not done). Bold: best method; underlined: best structural predictor. On A15 and MOOC, which have no containers, CONT reduces to ORDER; ZPD and DKT need recorded outcomes, which the MOOC log lacks.}
\label{tab:main}
\small
\setlength{\tabcolsep}{5pt}
\resizebox{\linewidth}{!}{\begin{tabular}{lrrrrrrrr}
\toprule
Method & A09 & A12 & A15 & A17 & EdNet & Junyi & Eedi & MOOC \\
\midrule
\multicolumn{9}{l}{\emph{Reference}} \\
RAND & 0.000 & 0.000 & 0.050 & 0.002 & 0.001 & 0.007 & 0.065 & 0.071 \\
\multicolumn{9}{l}{\emph{Structural (not personalised)}} \\
POP & 0.004 & 0.002 & 0.105 & 0.008 & 0.009 & 0.147 & 0.189 & 0.243 \\
ORDER & 0.002 & 0.005 & 0.059 & 0.043 & 0.001 & 0.069 & 0.277 & 0.457 \\
CONT & \underline{\textbf{0.083}} & 0.081 & 0.059 & 0.253 & 0.010 & 0.293 & 0.340 & 0.457 \\
MC & 0.083 & \underline{\textbf{0.276}} & \underline{0.420} & \underline{0.713} & \underline{0.035} & \underline{0.488} & \underline{0.558} & \underline{0.692} \\
\multicolumn{9}{l}{\emph{Personalised}} \\
ItemKNN & 0.078 & 0.222 & 0.408 & 0.617 & 0.040 & 0.421 & 0.470 & 0.503 \\
RP3beta & 0.049 & 0.158 & 0.396 & 0.203 & 0.024 & 0.376 & 0.280 & 0.345 \\
EASE & 0.050 & 0.095 & 0.470 & 0.374 & 0.024 & 0.432 & 0.401 & 0.515 \\
iALS & 0.033 & 0.064 & 0.319 & 0.074 & 0.016 & 0.389 & 0.403 & 0.504 \\
SKNN & 0.065 & 0.195 & 0.496 & 0.520 & 0.037 & 0.434 & 0.521 & 0.546 \\
MCLD & 0.039 & 0.042 & 0.447 & 0.006 & 0.001 & 0.256 & 0.266 & 0.376 \\
aMCLD & 0.037 & 0.085 & 0.433 & 0.034 & 0.008 & 0.286 & 0.262 & 0.421 \\
ZPD & 0.003 & 0.001 & 0.066 & 0.003 & 0.001 & 0.089 & 0.096 & -- \\
DKT & 0.001 & 0.000 & 0.057 & 0.009 & 0.001 & 0.015 & 0.051 & -- \\
SASRec & 0.077 & 0.256 & \textbf{0.522} & \textbf{0.728} & \textbf{0.042} & \textbf{0.535} & \textbf{0.559} & \textbf{0.780} \\
\bottomrule
\end{tabular}
}
\end{table}
\begin{table}[t]
\centering
\caption{Lift over structure under protocol P1 (NDCG@10). $\Delta$: best personalised recommender minus best structural predictor, with a paired bootstrap 95\% interval; $p_{\text{Holm}}$: bootstrap $p$-value of $\Delta$ with Holm's correction across datasets; $\rho$: structural share. $\rho_{\text{meta}}$: the same share for the structure without population transitions (best of POP, ORDER and CONT).}
\label{tab:lift}
\small
\resizebox{\linewidth}{!}{\begin{tabular}{llrlrrrrr}
\toprule
Dataset & Best structural & $S_{\text{struct}}$ & Best personalised & Score & $\Delta$ [95\% interval] & $p_{\text{Holm}}$ & $\rho$ & $\rho_{\text{meta}}$ \\
\midrule
A09 & CONT & 0.083 & ItemKNN & 0.078 & -0.005 [-0.013, +0.003] & 0.360 & 1.00 & 1.00 \\
A12 & MC & 0.276 & SASRec & 0.256 & -0.020 [-0.023, -0.017] & $<$0.001 & 1.00 & 0.29 \\
A15 & MC & 0.420 & SASRec & 0.522 & +0.102 [+0.096, +0.108] & $<$0.001 & 0.81 & 0.20 \\
A17 & MC & 0.713 & SASRec & 0.728 & +0.015 [+0.008, +0.022] & $<$0.001 & 0.98 & 0.35 \\
EdNet & MC & 0.035 & SASRec & 0.042 & +0.007 [+0.003, +0.012] & $<$0.001 & 0.82 & 0.23 \\
Junyi & MC & 0.488 & SASRec & 0.535 & +0.047 [+0.044, +0.051] & $<$0.001 & 0.91 & 0.55 \\
Eedi & MC & 0.558 & SASRec & 0.559 & +0.001 [-0.006, +0.008] & 0.779 & 1.00 & 0.61 \\
MOOC & MC & 0.692 & SASRec & 0.780 & +0.088 [+0.083, +0.093] & $<$0.001 & 0.89 & 0.59 \\
\bottomrule
\end{tabular}
}
\end{table}
Table~\ref{tab:main} gives the scores of all methods, and Table~\ref{tab:lift} summarises the comparison between structure and personalisation (Figure~\ref{fig:main} in Appendix~\ref{app:more} shows it as bars). \textbf{MC is the best structural predictor on \numnStructMC{} of the \numnDatasets{} datasets} (CONT on A09). \textbf{The structural share $\rho$ ranges from \numrhoMin{} to \numrhoMax{}, with a median of \numrhoMedian{}}: on most datasets, a predictor that knows nothing about the learner except the last activity already delivers nearly all of the best offline accuracy that any of the ten personalised methods reaches. The best personalised recommender is significantly better than the best structural predictor on \numsigList{} (Holm-adjusted paired bootstrap, $\alpha=0.05$), significantly worse on \numworseList{}, and not significantly different on \numnsList{}. The largest lifts appear on A15 ($\Delta=\numdeltaAfifteen{}$), where activities are whole problem sets and the catalogue is small, on the MOOC ($\Delta=\numdeltaMOOC{}$) and on Junyi ($\Delta=\numdeltaJunyi{}$). The significant lifts on A17 ($\Delta=\numdeltaAseventeen{}$) and EdNet ($\Delta=\numdeltaEdNet{}$) are small in absolute terms, although on EdNet, where every method is weak, the latter is about a fifth of the structural score. All five come from SASRec, which reached its training budget on \numsasCapList{} (Appendix~\ref{app:hparams}); its scores there are lower bounds. Three further trainings of SASRec with other seeds on \numseedN{} datasets (Appendix~\ref{app:seeds}) confirm these conclusions on \numseedStableList{}, where every training gives a significant lift and SASRec's NDCG@10 varies by a standard deviation of at most \numseedSdStableMax{}. On Eedi they do not: the three other trainings give significant lifts of \numseedDminExtraEedi{} to \numseedDmaxExtraEedi{}, so there the conclusion depends on the training run. Choosing the best structural and personalised methods by their validation scores instead of their test scores changes the choice on \numvalChanged{} dataset and none of these conclusions (Appendix~\ref{app:valsel}).

\paragraph{Syllabus or paths?} The structural share comes mostly from the population transitions, not from the course metadata. Without MC, the best of POP, ORDER and CONT reaches between \numrhoMetaMin{} and \numrhoMetaMax{} of the best accuracy (median \numrhoMetaMedian{}; Table~\ref{tab:lift}, Figure~\ref{fig:teaser}). It exceeds half of it only on A09, where the container (the problem set from which the tutor draws problems) is the best structural predictor, and on Junyi, Eedi and the MOOC (\numrhoMetaJunyi{}, \numrhoMetaEedi{} and \numrhoMetaMOOC{}). The shortcut is therefore the typical paths that assignments and engines produce, which the logs record, more than the syllabus as written.

\paragraph{Which personalised methods.} SASRec, which conditions on the order of the history, is the best personalised method on \numnSASRecBest{} of the \numnDatasets{} datasets. SKNN, which also looks at the most recent activities, is second on A15, Junyi, Eedi and the MOOC, but beats MC only on A15 and, marginally, on EdNet (\numvEdNetSKNN{} against \numvEdNetMC{}). The static collaborative filters (EASE, RP3beta, iALS) and the peer-based methods, which use the set of past activities rather than its order, fall below MC on the datasets with strongly sequenced logs (far below on A17 and Eedi) and beat it only on A15. MCLD, designed for small classes following the same course, is most competitive on A15 (\numvAfifteenMCLD{} against \numvAfifteenSASRec{} for SASRec), where problem sets play the role of course units. The knowledge-tracing policies rank below popularity: ZPD on all \numnZPDbelowPOP{} datasets with recorded outcomes, and DKT on \numnDKTbelowPOP{} of them, although DKT predicts the outcome of the target step better than ZPD's smoothed rates on all of them (Section~\ref{sec:methods}). A better model of the learner does not help a policy that does not try to imitate the logs: recommending activities at a desirable difficulty disagrees with what the platform made learners do next, which offline next-activity evaluation can only score as an error.

\subsection{RQ3: Where does personalisation help?}\label{sec:rq3}
\begin{table}[t]
\centering
\caption{Seen and novel transitions (protocol P1, NDCG@10). A test step is \emph{novel} when its activity never followed the learner's last activity in the training data. \emph{Novel}: share of test steps. \emph{Struct.} and \emph{Pers.}: best structural predictor and best personalised recommender within each group; $\Delta$: their difference on novel steps. This definition needs no container.}
\label{tab:novel}
\small
\setlength{\tabcolsep}{4pt}
\resizebox{\linewidth}{!}{\begin{tabular}{lrrrrrlr}
\toprule
 & & \multicolumn{2}{c}{Seen transitions} & \multicolumn{4}{c}{Novel transitions} \\
\cmidrule(lr){3-4}\cmidrule(lr){5-8}
Dataset & Novel & Struct. & Pers. & Struct. & Pers. & Best pers. & $\Delta$ novel \\
\midrule
A09 & 72\% & 0.298 & 0.251 & 0.053 & 0.031 & ItemKNN & -0.022 \\
A12 & 43\% & 0.484 & 0.439 & 0.024 & 0.033 & ItemKNN & +0.009 \\
A15 & 4\% & 0.436 & 0.539 & 0.037 & 0.162 & MCLD & +0.125 \\
A17 & 15\% & 0.841 & 0.854 & 0.017 & 0.045 & ItemKNN & +0.028 \\
EdNet & 66\% & 0.103 & 0.100 & 0.005 & 0.013 & SASRec & +0.008 \\
Junyi & 10\% & 0.541 & 0.591 & 0.016 & 0.064 & RP3beta & +0.048 \\
Eedi & 4\% & 0.580 & 0.580 & \multicolumn{4}{c}{(fewer than 200 cases)} \\
MOOC & 1\% & 0.698 & 0.784 & \multicolumn{4}{c}{(fewer than 200 cases)} \\
\bottomrule
\end{tabular}
}
\end{table}
\paragraph{Novel transitions.} A definition of hard steps that needs no container is a step whose activity never followed the last activity in the training data (Table~\ref{tab:novel}). Such steps make up \numnovshareAnine{}\% of the test steps of A09, \numnovshareEdNet{}\% of EdNet and \numnovshareAtwelve{}\% of A12, but at most \numnovshareAseventeen{}\% elsewhere. On the \numnovN{} datasets with at least 200 such steps, every method does poorly on them: the best structural predictor reaches an NDCG@10 of at most \numnovSmax{} and the best personalised recommender at most \numnovPmax{} (\numnovPmaxWhere{}), against \numseenSmin{} to \numseenSmax{} for the best structural predictor on seen transitions. Personalisation gains something on novel steps on A15 and Junyi ($\Delta=\numDnovAfifteen{}$ and \numDnovJunyi{}), but its accuracy there stays far below its accuracy on seen transitions. On A09 and EdNet, where most test steps are novel, this accounts for much of the low overall scores of every method.

\figpaths
\paragraph{Leaving the container.} On four of the six datasets with containers, off-path steps, whose activity is outside the container of the last one, are much harder than on-path steps for both the best structural and the best personalised method (Figure~\ref{fig:paths}; Table~\ref{tab:paths} in Appendix~\ref{app:more}). The best structural predictor reaches NDCG@10 of \numSonAseventeen{} on on-path steps and \numSoffAseventeen{} on off-path steps on A17, \numSonJunyi{} against \numSoffJunyi{} on Junyi, \numSonEedi{} against \numSoffEedi{} on Eedi and \numSonAnine{} against \numSoffAnine{} on A09. On A12, off-path steps are predicted better than on-path ones (\numSoffAtwelve{} against \numSonAtwelve{}), and on EdNet the two kinds of steps are about equally hard. The lift of personalisation is larger on off-path than on on-path steps on A17 (\numDoffAseventeen{} against \numDonAseventeen{}) and Eedi (\numDoffEedi{} against \numDonEedi{}), which is consistent with personalisation mattering where a choice is made, but not on Junyi (\numDoffJunyi{} against \numDonJunyi{}), and on EdNet the two are about equal (\numDoffEdNet{} against \numDonEdNet{}). In no case does it close the gap: on the four datasets where off-path steps are harder, the best personalised method stays far below its on-path accuracy.

\begin{table}[t]
\centering
\caption{NDCG@10 by who started the target activity (protocol P1). $\rho$: best structural score divided by the best score within the group.}
\label{tab:sources}
\small
\resizebox{\linewidth}{!}{\begin{tabular}{llrrrrrr}
\toprule
Dataset & Who started the target activity & Cases & MC & CONT & Best pers. & SASRec & $\rho$ \\
\midrule
EdNet & app chose the set (sprint) & 2,900 & 0.025 & 0.010 & 0.035 & 0.035 & 0.71 \\
EdNet & app initiated (offer, tutor, diagnosis) & 858 & 0.066 & 0.003 & 0.069 & 0.069 & 0.86 \\
EdNet & review modes & 609 & 0.040 & 0.015 & 0.041 & 0.041 & 0.95 \\
\midrule
Junyi & platform suggestion (map) & 2,661 & 0.649 & 0.466 & 0.726 & 0.726 & 0.89 \\
Junyi & opened from a topic & 5,349 & 0.658 & 0.353 & 0.714 & 0.714 & 0.92 \\
Junyi & chosen by the learner & 5,996 & 0.264 & 0.162 & 0.292 & 0.292 & 0.91 \\
\bottomrule
\end{tabular}
}
\end{table}
\paragraph{Who started the activity.} The logs of EdNet and Junyi tell who started each activity (Table~\ref{tab:sources}). On Junyi, exercises suggested by the platform from its prerequisite map or opened from a topic are predicted far better than exercises chosen by the learner, by structural and personalised methods alike: the best structural predictor reaches \numsrcSJunyiplatformsuggestionmap{} and \numsrcSJunyiopenedfromatopic{} on the first two groups and \numsrcSJunyichosenbythelearner{} on learner-chosen exercises. Personalised methods do not close this gap. Their lift over the structure is \numsrcLiftJunyichosenbythelearner{} on learner-chosen exercises, against \numsrcLiftJunyiplatformsuggestionmap{} on the platform's suggestions, which are themselves computed from the learner's proficiency; relative to the structural score, the lift is similar in the three groups (structural share \numsrcJunyiplatformsuggestionmap{} to \numsrcJunyiopenedfromatopic{}). Learned recommenders thus gain no more where the learner exercised a choice than where they can imitate the platform's suggestions. On EdNet, where the application selects the question sets in almost all modes, every method is weak in absolute terms (the catalogue has \numitemsEdNet{} activities), and the structural share ranges from \numsrcEdNetappchosethesetsprint{} to \numsrcEdNetreviewmodes{} across groups.

\subsection{RQ4: LLM rankers and the choice of negatives}\label{sec:rq4}
\begin{table}[t]
\centering
\caption{HR@1 on candidate sets of 20 activities (the target and 19 negatives the learner has not done), with random or structural negatives; 200 test cases per dataset and type. RAND is the expected value of a random ranking. LEX ranks candidates by word overlap between their name and the last activity's name. The LLMs score candidates by likelihood (Section~\ref{sec:llmsetup}). Bold: best value in the column.}
\label{tab:llm}
\scriptsize
\setlength{\tabcolsep}{4pt}
\begin{tabular}{lrrrrrrrr}
\toprule
 & \multicolumn{2}{c}{A09} & \multicolumn{2}{c}{EdNet} & \multicolumn{2}{c}{Junyi} & \multicolumn{2}{c}{Eedi} \\
\cmidrule(lr){2-3}\cmidrule(lr){4-5}\cmidrule(lr){6-7}\cmidrule(lr){8-9}
Method & random & struct. & random & struct. & random & struct. & random & struct. \\
\midrule
RAND & 0.050 & 0.050 & 0.050 & 0.050 & 0.050 & 0.050 & 0.050 & 0.050 \\
POP & 0.158 & 0.010 & 0.205 & 0.060 & 0.355 & 0.105 & 0.110 & 0.055 \\
ORDER & 0.135 & 0.025 & 0.115 & 0.000 & 0.217 & 0.100 & 0.210 & 0.172 \\
CONT & \textbf{0.650} & 0.065 & 0.255 & 0.005 & 0.495 & 0.255 & 0.253 & 0.175 \\
MC & 0.323 & 0.040 & 0.245 & 0.015 & 0.745 & 0.430 & 0.460 & 0.285 \\
ItemKNN & 0.595 & \textbf{0.110} & 0.320 & 0.060 & 0.660 & 0.325 & 0.415 & 0.305 \\
EASE & 0.330 & 0.085 & 0.305 & \textbf{0.120} & 0.740 & 0.390 & 0.350 & 0.255 \\
SKNN & 0.540 & 0.052 & 0.335 & 0.060 & 0.718 & 0.370 & \textbf{0.515} & \textbf{0.355} \\
MCLD & 0.190 & 0.070 & 0.140 & 0.083 & 0.537 & 0.215 & 0.225 & 0.125 \\
SASRec & 0.590 & 0.050 & \textbf{0.445} & 0.090 & \textbf{0.795} & \textbf{0.465} & 0.470 & 0.310 \\
LEX & 0.594 & 0.064 & 0.233 & 0.039 & 0.568 & 0.177 & 0.172 & 0.148 \\
\midrule
Qwen2.5-0.5B & 0.600 & 0.095 & 0.190 & 0.055 & 0.410 & 0.150 & 0.175 & 0.155 \\
Qwen2.5-1.5B & \textbf{0.650} & \textbf{0.110} & 0.210 & 0.055 & 0.420 & 0.165 & 0.180 & 0.155 \\
\bottomrule
\end{tabular}

\end{table}
\figllm
Table~\ref{tab:llm} and Figure~\ref{fig:llm} compare all methods on the same candidate sets. With random negatives the LLMs can look competent: on A09 the 0.5B model places the target first in \numllmAnineRandQwenSmall{} of the cases, close to container continuation (\numllmAnineRandCONT{}). But LEX, which only counts the words that a candidate's name shares with the last activity's name, does as well (\numllmAnineRandLEX{}). Over the four datasets, the better LLM exceeds LEX by at most \numllmMaxOverLex{} with random negatives. With random negatives the target is usually the only candidate on the skill or topic of the last activity, and recognising this suffices. With structural negatives, which are all plausible continuations, the LLMs fall to \numllmAnineHardQwenSmall{} (0.5B) and \numllmAnineHardQwenLarge{} (1.5B) on A09 and to \numllmEdNetHardQwenSmall{} on EdNet, against a random level of 0.05, and every other method also loses most of its accuracy: on these two datasets no method exceeds \numllmBestEdNetHard{}. Popularity falls to \numllmAnineHardPOP{} on A09 and \numllmEediHardPOP{} on Eedi, because the structural negatives are themselves popular activities. Where some signal survives structural negatives (Junyi and Eedi), the methods that learned the platform's paths (SASRec, SKNN and MC) keep a large lead over the LLMs. %
\begin{table}[t]
\centering
\caption{Model size. HR@1 on the first \numscaleN{} cases of each candidate set, for all four Qwen2.5 sizes and three reference methods (rand.: random negatives; struct.: structural negatives; random level 0.05). The 3B and 7B models are 4-bit quantized. Bold: best in the column.}
\label{tab:scale}
\small
\setlength{\tabcolsep}{4pt}
\begin{tabular}{lrrrrrrrr}
\toprule
 & \multicolumn{2}{c}{A09} & \multicolumn{2}{c}{EdNet} & \multicolumn{2}{c}{Junyi} & \multicolumn{2}{c}{Eedi} \\
 & rand. & struct. & rand. & struct. & rand. & struct. & rand. & struct. \\
\midrule
LEX & \textbf{0.64} & 0.07 & 0.24 & 0.04 & 0.56 & 0.19 & 0.16 & 0.12 \\
MC & 0.32 & 0.03 & 0.31 & 0.01 & 0.80 & 0.46 & 0.45 & 0.30 \\
SASRec & 0.54 & 0.03 & \textbf{0.47} & \textbf{0.11} & \textbf{0.85} & \textbf{0.50} & \textbf{0.47} & \textbf{0.32} \\
\midrule
Qwen2.5 0.5B & 0.56 & \textbf{0.09} & 0.18 & 0.06 & 0.44 & 0.14 & 0.17 & 0.15 \\
Qwen2.5 1.5B & 0.62 & 0.08 & 0.22 & 0.05 & 0.37 & 0.15 & 0.18 & 0.14 \\
Qwen2.5 3B (4-bit) & 0.58 & 0.07 & 0.21 & 0.04 & 0.36 & 0.12 & 0.21 & 0.13 \\
Qwen2.5 7B (4-bit) & 0.59 & 0.06 & 0.24 & 0.05 & 0.43 & 0.12 & 0.15 & 0.16 \\
\bottomrule
\end{tabular}

\end{table}
Scale does not change this picture (Table~\ref{tab:scale}). On the same \numscaleN{} cases per set, the 7B model's HR@1 differs from the 0.5B model's by \numscaleGainMin{} to \numscaleGainMax{}; with random negatives it beats LEX \numscaleSevenBeatsLex{} of the four datasets; with structural negatives no model size exceeds \numscaleLLMmaxHardAnineEdNet{} on A09 and EdNet, and on Junyi and Eedi MC leads the best LLM by at least \numscaleMCleadMin{}. Within this range, a fourteen-fold increase in parameters brings no systematic gain on this task. Results of LLM rankers on random candidate sets are therefore uninterpretable without structural predictors and LEX scored on the same sets: matching the topic of the last activity can be mistaken for recommendation ability.

\subsection{RQ5: Robustness to the evaluation protocol}\label{sec:rq5}
Under the global temporal split (Figure~\ref{fig:teaser}; Table~\ref{tab:p2} and all methods in Appendix~\ref{app:p2}), absolute scores are lower: the best structural score is between \numPtwoSratioMin{} and \numPtwoSratioMax{} times its value under P1. Here the models see only the period before the cut-off, and each test case is a learner's first new activity after it. The comparison between structure and personalisation is otherwise similar. MC is again the best structural predictor on \numPtwonStructMC{} of the \numnDatasets{} datasets (CONT on A09), and the best personalised recommender is significantly better on \numPtwosigList{}, significantly worse on \numPtwoworseList{}, and not significantly different on \numPtwonsList{}; on A09 the test set contains only \numPtwoAninenTest{} learners. The structural share ranges from \numPtworhoMin{} to \numPtworhoMax{} (median \numPtworhoMedian{}). It is lower than under P1 on A15, where the peer-based MCLD is the best method (\numPtwoAfifteenP{} against \numPtwoAfifteenS{} for MC), and on EdNet ($\rho=\numPtwoEdNetrho{}$), where the absolute scores are tiny. Under both protocols the structure delivers most of the attainable accuracy on most datasets.

\section{What offline accuracy misses: a check with known needs}\label{sec:sim}
The real logs cannot tell which recommendation a learner needed. To check what the summary quantities and offline accuracy measure when the answer is known, we simulate logs in which the share of steps decided by the platform is a parameter.

\paragraph{Simulation.} A course has 10 units of 10 activities in a fixed order. Each of 3,000 learners has three weak units, where they succeed 45\% of the time (85\% elsewhere), and logs 20 to 40 new activities. At each step, with probability $\alpha$ the next activity follows the platform path: the next unseen activity in course order, or, with probability 0.2, another unseen activity of the current unit. Otherwise the learner works on their needs: an unseen activity of a weak unit, at random. All methods of Section~\ref{sec:methods} are run unchanged under protocol P1 (SASRec with a 10-minute budget). Because the needs are known, every method also gets a score of \emph{true usefulness}, need@10: the share of its top 10 that are unseen activities of the learner's weak units. Two oracles complete the comparison: NEED recommends exactly the unseen activities of the weak units, and PI ranks by the true probabilities of the logging policy.

\begin{table}[t]
\centering
\caption{Simulation with known needs (Section~\ref{sec:sim}). $\alpha$: share of steps decided by the platform path; $\rho$: structural share; need@10: share of a method's top 10 that are unseen activities of the learner's weak units (true usefulness); NEED: oracle that recommends exactly these activities. Kendall $\tau$ over the 11 methods between true usefulness and plain or structure-corrected NDCG@10, and the rank of NEED among them by each.}
\label{tab:sim}
\small
\setlength{\tabcolsep}{3.5pt}
\resizebox{\linewidth}{!}{\begin{tabular}{rrrrrrrrrrr}
\toprule
 & & \multicolumn{3}{c}{Offline NDCG@10} & \multicolumn{3}{c}{True usefulness (need@10)} & \multicolumn{2}{c}{Kendall $\tau$ with need@10} & Rank of NEED \\
\cmidrule(lr){3-5}\cmidrule(lr){6-8}\cmidrule(lr){9-10}
$\alpha$ & $\rho$ & MC & SASRec & NEED & MC & SASRec & NEED & NDCG & corrected & NDCG / corr. \\
\midrule
0.00 & 0.53 & 0.190 & 0.360 & 0.359 & 0.39 & 1.00 & 1.00 & 0.86 & 0.75 & 3 / 1 \\
0.25 & 0.77 & 0.351 & 0.453 & 0.361 & 0.39 & 0.95 & 1.00 & 0.71 & 0.85 & 5 / 2 \\
0.50 & 0.90 & 0.514 & 0.573 & 0.357 & 0.39 & 0.77 & 1.00 & 0.42 & 0.85 & 7 / 2 \\
0.75 & 0.99 & 0.698 & 0.705 & 0.268 & 0.37 & 0.40 & 1.00 & 0.24 & 0.75 & 7 / 2 \\
0.90 & 0.99 & 0.815 & 0.819 & 0.166 & 0.34 & 0.35 & 1.00 & 0.16 & 0.35 & 10 / 7 \\
\bottomrule
\end{tabular}
}
\end{table}
\figsim
\paragraph{Results.} Table~\ref{tab:sim} and Figure~\ref{fig:sim} show three things. First, the structural share tracks the platform's share of the decisions: $\rho$ rises from \numsimRhoAAA{} when learners choose every step to \numsimRhoAHF{} when the platform decides three steps out of four. Second, offline accuracy stops measuring usefulness as $\alpha$ grows. NEED, which by construction recommends exactly what each learner needs, ranks \numsimRankNeedPlainAAA{}rd of 11 methods by NDCG@10 at $\alpha=0$ and \numsimRankNeedPlainAHF{}th at $\alpha=0.75$, while SASRec and MC, with a need@10 of only \numsimSASRecNeedAHF{} and \numsimMCNeedAHF{}, come first and second. The rank correlation between NDCG@10 and true usefulness over the 11 methods falls from \numsimTauPlainAAA{} to \numsimTauPlainAJA{} at $\alpha=0.9$. Third, learned recommenders follow the logs: SASRec is as useful as NEED when learners decide (need@10 of \numsimSASRecNeedAAA{}) and no more useful than MC when the platform does (\numsimSASRecNeedAJA{} at $\alpha=0.9$), while its offline accuracy keeps rising. A recommender tuned and selected by offline accuracy on platform-driven logs thus learns the platform.

\paragraph{Structure-corrected accuracy.} One way to look past the platform is to weight each test step by the inverse of its structural propensity: $w=1/\max(p,0.01)$, where $p$ is the share of training learners who went from the step's last activity to its target (the transition probability of MC), and to report the weighted mean of NDCG@10, a self-normalised inverse-propensity estimate in the sense of \citet{schnabel2016}. Steps that the population's paths make likely then count less and unlikely ones more; the floor bounds every weight at 100. This is not an estimate of online performance: it asks how well a method predicts the steps that the structure does not explain. In the simulation it restores much of the link with usefulness while the learners still decide a quarter of the steps: its rank correlation with need@10 is \numsimTauCorrAFA{} at $\alpha=0.5$ and \numsimTauCorrAHF{} at $\alpha=0.75$ (NDCG@10: \numsimTauPlainAFA{} and \numsimTauPlainAHF{}), and it ranks NEED second at both. When the platform decides nine steps out of ten, little is left to weight and it fails too (\numsimTauCorrAJA{}).

\begin{table}[t]
\centering
\caption{Structure-corrected accuracy (protocol P1): NDCG@10 with each test step weighted by the inverse of its transition probability under MC (floored at 0.01), so that steps the structure does not explain count more. $\Delta_{\text{corr}}$: best personalised minus best structural, with a paired bootstrap 95\% interval; $\rho_{\text{corr}}$: corrected structural share.}
\label{tab:corrected}
\small
\resizebox{\linewidth}{!}{\begin{tabular}{llrlrrr}
\toprule
Dataset & Best structural & Score & Best personalised & Score & $\Delta_{\text{corr}}$ [95\% interval] & $\rho_{\text{corr}}$ \\
\midrule
A09 & CONT & 0.062 & ItemKNN & 0.045 & -0.017 [-0.024, -0.010] & 1.00 \\
A12 & MC & 0.044 & ItemKNN & 0.066 & +0.022 [+0.020, +0.025] & 0.66 \\
A15 & MC & 0.121 & MCLD & 0.367 & +0.245 [+0.236, +0.255] & 0.33 \\
A17 & MC & 0.122 & SASRec & 0.161 & +0.039 [+0.025, +0.054] & 0.76 \\
EdNet & MC & 0.023 & SASRec & 0.034 & +0.012 [+0.008, +0.016] & 0.66 \\
Junyi & MC & 0.119 & SASRec & 0.192 & +0.074 [+0.069, +0.078] & 0.62 \\
Eedi & MC & 0.223 & SKNN & 0.328 & +0.105 [+0.094, +0.117] & 0.68 \\
MOOC & MC & 0.395 & SASRec & 0.597 & +0.202 [+0.187, +0.217] & 0.66 \\
\bottomrule
\end{tabular}
}
\end{table}
On the real datasets (Table~\ref{tab:corrected}), the correction changes the picture. The best personalised recommender beats the best structural predictor on the steps the structure does not explain on \numcorrSigN{} of the \numnDatasets{} datasets (\numcorrSigList{}; bootstrap 95\% intervals), all but A09, and the structural share falls to between \numrhoCorrMin{} and \numrhoCorrMax{} (median \numrhoCorrMedian{}). The largest corrected lifts appear on A15, where the peer-based MCLD is the best method ($\Delta_{\text{corr}}=\numdCorrAfifteen{}$), and on the MOOC ($\Delta_{\text{corr}}=\numdCorrMOOC{}$). Personalisation thus carries a real signal on the steps the structure does not explain; plain NDCG@10 buries it under the steps the structure does. Because the weights come from MC itself, this view is deliberately unfavourable to MC, and we report it next to plain accuracy, not instead of it.

\section{Discussion}\label{sec:discussion}

\paragraph{What offline next-activity accuracy measures.}
On public educational logs, next-activity accuracy is dominated by how well a method reproduces the sequencing that teachers and platforms imposed. This does not make offline evaluation useless, but it changes what a result means. A method that beats the structural predictors by a clear margin predicts logged choices better than the platform's typical paths; a method that does not, whatever its absolute score, has shown nothing about personalisation. Conversely, a policy designed for learning, such as practising at a desirable difficulty, is penalised precisely because it departs from the logged policy, so offline accuracy can rank pedagogically motivated recommenders below simple imitators. The simulation of Section~\ref{sec:sim} shows how far this can go: when the platform decides most steps, a recommender that serves every learner's needs ranks near the bottom by offline accuracy, and a recommender selected by offline accuracy learns the platform. The consequence for the field is not to stop offline evaluation but to report it against the right floor and on the right steps.

\paragraph{Why the shortcut is strong in education.}
Educational platforms are designed to sequence. Mastery-based problem sets draw the next problem from a small pool (A09), applications serve questions part by part (EdNet), knowledge maps suggest the next exercise from prerequisites (Junyi), teachers assign quizzes within schemes of work (Eedi), and a MOOC lays out its activities in a course. Each of these mechanisms produces low-branching, highly repeatable transitions, the very structures that make sequential benchmarks shortcut-solvable \citep{han2026shortcut}. In addition, the datasets most used for educational recommendation were collected for knowledge tracing, a task for which the sequencing was irrelevant; they were never meant to evaluate the choice of activities.

\paragraph{Implications for peer-based and LLM-based recommenders.}
Peer-based methods such as MCLD were designed for a class following one course, with recommendations that come with a one-sentence explanation \citep{jdidou2016,jdidou2021}. Next-new-activity accuracy rewards neither their transparency nor their remediation component, and on long, strongly sequenced histories their set-based view of the learner is outperformed by the transition heuristic. On A15, where activities are whole problem sets, MCLD is close to the best methods under leave-last-out (\numvAfifteenMCLD{} against \numvAfifteenSASRec{} for SASRec) and the best method under the global temporal split. Their value has to be established on the steps where a choice is made, and ideally in deployment. For LLM-based recommenders, our results with open models of up to 7 billion parameters are a caution rather than a verdict: much larger models may well rank candidates better than the ones we could run, although within our range size made no difference. Whatever the model, its accuracy on candidate sets is uninterpretable unless the structural predictors and a simple lexical baseline are scored on the same sets, and unless structural negatives are reported.

\subsection{A syllabus-aware protocol for offline evaluation}\label{sec:protocolrec}
The results suggest a short checklist for anyone reporting offline results of an educational recommender. Every item is cheap; the accompanying package implements the first three in a few lines of code.
\begin{enumerate}[leftmargin=*,itemsep=2pt]
\item \textbf{Report the structural predictors.} Include popularity, the course order (ORDER), container continuation (CONT) and population transitions (MC), evaluated on exactly the same cases and candidates as the proposed method. They need no training and no hyperparameters, so there is no reason to omit them.
\item \textbf{Report the lift over structure, not only the metric.} Give $S_{\text{struct}}$, the lift $\Delta$ of each method over it with a confidence interval, and the structural share $\rho$. A raw NDCG@10 of 0.7 means little if the course order alone reaches 0.7. Add the structure-corrected accuracy of Section~\ref{sec:sim}, which weights each test step by the inverse of its transition probability, to show how a method does on the steps the structure does not explain.
\item \textbf{Split by kind of step.} Report the share and the metrics of steps along transitions never seen in training and, where containers exist, of off-path steps: this is where a recommender has a choice to make. When the logs record who initiated each activity, split by initiator as well.
\item \textbf{Choose negatives deliberately in sampled evaluation.} For LLM rankers and any evaluation on candidate sets, report results with structural negatives (candidates that the course structure finds plausible) in addition to random ones, and always score the structural predictors on the same sets.
\item \textbf{Prefer temporal splits and state the protocol.} Leave-last-out and a global temporal split agreed here on the best structural predictor and on most comparisons, but absolute scores were far lower under the temporal split and the structural share fell on two datasets. The temporal split also avoids training on the future of other learners \citep{ji2023leakage}.
\item \textbf{Do not read agreement with the logs as pedagogical value.} Next-activity accuracy measures how well a model imitates the sequencing policy that produced the data. Claims about learning require outcomes: randomised sequencing data, field experiments, or at least policy evaluation under several student models \citep{doroudi2017rem,doroudi2019reward}.
\end{enumerate}

\subsection{Limitations}\label{sec:limits}
The study is offline and observational: it shows how much of the measured accuracy is attainable from the structure, not which recommendations would help learners. The container information differs across datasets (problem sets, skills, parts, topics, schemes of work), and in A12 and A17 the skill stands in for the problem set, which probably understates the structural share there. The LLMs are open-weight models of up to 7 billion parameters run on CPUs, the two largest 4-bit quantized and scored on \numscaleN{} cases per set, chosen so that the experiment can be reproduced without special hardware or paid services; much larger models may behave differently, although the dependence of every method on the type of negatives does not depend on the LLM. The MOOC activities are anonymised, so the MOOC has no container or course metadata and cannot enter the LLM comparison, and DKT was trained with fixed settings and a short budget. EdNet is represented by a public sample of a few thousand learners and Junyi by a random sample of 20,000 learners. SASRec reached its training budget on \numsasCapList{}, its main results come from one training per dataset (four trainings on \numseedN{} datasets agree except on Eedi), and under P2 the A09 test set is small (\numPtwoAninenTest{} learners). Finally, the hyperparameter grids are modest; longer training and better-tuned personalised models could reduce $\rho$ somewhat, but they would have to close the large gaps observed on off-path and learner-chosen steps to change the conclusions.

\section{Conclusion}\label{sec:conclusion}
Recommenders and LLMs evaluated on educational logs largely learn the platform: they beat the syllabus, or more exactly the typical paths that teachers and platforms impose, by much less than their absolute scores suggest. Across eight datasets from five platforms, a predictor that only knows the last activity delivers between \numrhoMin{} and \numrhoMax{} of the best offline accuracy, mostly through the population's typical paths. Personalised recommenders improve on it significantly on \numnSig{} datasets, by \numsigDeltaMin{} to \numsigDeltaMax{} NDCG@10, but none does well on steps along unseen transitions where these are frequent enough to measure, and none closes the gap on steps that leave the current container or that learners choose themselves, and small open LLMs do not beat the best structural predictor on the same candidates except where every method is close to the random level. We hope that the summary quantities, the checklist and the code released with this paper help future work show what an educational recommender adds beyond the platform it was trained on.

\section*{Data and code availability}
All datasets are public and were obtained from the sources listed in the accompanying code. No learner data are redistributed. The code, the evaluation protocols and the per-case results needed to recompute every table are released with this preprint as ancillary files. The installable package \texttt{beat-the-syllabus} and all the scripts of the study are available at \url{https://github.com/jdidou/beat-the-syllabus} and archived on Zenodo \citep{jdidou2026code}, \href{https://doi.org/10.5281/zenodo.24443552}{doi:10.5281/zenodo.24443552}.

\section*{Ethics statement}
The study uses only previously published learner logs in which learners are identified by anonymous codes. No data were collected from people for this study, and no attempt was made to identify any learner.

\section*{Use of generative AI}
During the preparation of this work, the author used Claude (Anthropic) for language editing, text restructuring, and stylistic refinement. Following its use, the author carefully reviewed and revised the manuscript, independently designed the study, verified the code, results, and references, and assumes full responsibility for the accuracy, integrity, and final content of the work.

\section*{Competing interests}
The author declares no competing interests. This research received no specific funding.

\bibliographystyle{plainnat}
\bibliography{refs}

@inproceedings{ferraridacrema2019,
  author    = {Ferrari Dacrema, Maurizio and Cremonesi, Paolo and Jannach, Dietmar},
  title     = {Are We Really Making Much Progress? A Worrying Analysis of Recent Neural Recommendation Approaches},
  booktitle = {Proceedings of the 13th ACM Conference on Recommender Systems},
  pages     = {101--109},
  year      = {2019},
  publisher = {ACM},
  doi       = {10.1145/3298689.3347058}
}

@article{ferraridacrema2021,
  author  = {Ferrari Dacrema, Maurizio and Boglio, Simone and Cremonesi, Paolo and Jannach, Dietmar},
  title   = {A Troubling Analysis of Reproducibility and Progress in Recommender Systems Research},
  journal = {ACM Transactions on Information Systems},
  volume  = {39},
  number  = {2},
  pages   = {1--49},
  year    = {2021},
  doi     = {10.1145/3434185}
}

@inproceedings{rendle2020ncf,
  author    = {Rendle, Steffen and Krichene, Walid and Zhang, Li and Anderson, John},
  title     = {Neural Collaborative Filtering vs. Matrix Factorization Revisited},
  booktitle = {Proceedings of the 14th ACM Conference on Recommender Systems},
  pages     = {240--248},
  year      = {2020},
  publisher = {ACM},
  doi       = {10.1145/3383313.3412488}
}

@inproceedings{krichene2020,
  author    = {Krichene, Walid and Rendle, Steffen},
  title     = {On Sampled Metrics for Item Recommendation},
  booktitle = {Proceedings of the 26th ACM SIGKDD International Conference on Knowledge Discovery \& Data Mining},
  pages     = {1748--1757},
  year      = {2020},
  publisher = {ACM},
  doi       = {10.1145/3394486.3403226}
}

@article{ji2023leakage,
  author  = {Ji, Yitong and Sun, Aixin and Zhang, Jie and Li, Chenliang},
  title   = {A Critical Study on Data Leakage in Recommender System Offline Evaluation},
  journal = {ACM Transactions on Information Systems},
  volume  = {41},
  number  = {3},
  pages   = {1--27},
  year    = {2023},
  doi     = {10.1145/3569930}
}

@inproceedings{hidasi2023flaws,
  author    = {Hidasi, Bal{\'a}zs and Czapp, {\'A}d{\'a}m Tibor},
  title     = {Widespread Flaws in Offline Evaluation of Recommender Systems},
  booktitle = {Proceedings of the 17th ACM Conference on Recommender Systems},
  pages     = {848--855},
  year      = {2023},
  publisher = {ACM},
  doi       = {10.1145/3604915.3608839}
}

@article{ludewig2021,
  author  = {Ludewig, Malte and Mauro, Noemi and Latifi, Sara and Jannach, Dietmar},
  title   = {Empirical Analysis of Session-Based Recommendation Algorithms: A Comparison of Neural and Non-Neural Approaches},
  journal = {User Modeling and User-Adapted Interaction},
  volume  = {31},
  number  = {1},
  pages   = {149--181},
  year    = {2021},
  doi     = {10.1007/s11257-020-09277-1}
}

@inproceedings{steck2019ease,
  author    = {Steck, Harald},
  title     = {Embarrassingly Shallow Autoencoders for Sparse Data},
  booktitle = {Proceedings of The World Wide Web Conference},
  pages     = {3251--3257},
  year      = {2019},
  publisher = {ACM},
  doi       = {10.1145/3308558.3313710}
}

@inproceedings{kang2018sasrec,
  author    = {Kang, Wang-Cheng and McAuley, Julian},
  title     = {Self-Attentive Sequential Recommendation},
  booktitle = {2018 IEEE International Conference on Data Mining (ICDM)},
  pages     = {197--206},
  year      = {2018},
  publisher = {IEEE},
  doi       = {10.1109/ICDM.2018.00035}
}

@inproceedings{hu2008ials,
  author    = {Hu, Yifan and Koren, Yehuda and Volinsky, Chris},
  title     = {Collaborative Filtering for Implicit Feedback Datasets},
  booktitle = {2008 Eighth IEEE International Conference on Data Mining},
  pages     = {263--272},
  year      = {2008},
  publisher = {IEEE},
  doi       = {10.1109/ICDM.2008.22}
}

@article{paudel2016rp3,
  author  = {Paudel, Bibek and Christoffel, Fabian and Newell, Chris and Bernstein, Abraham},
  title   = {Updatable, Accurate, Diverse, and Scalable Recommendations for Interactive Applications},
  journal = {ACM Transactions on Interactive Intelligent Systems},
  volume  = {7},
  number  = {1},
  pages   = {1--34},
  year    = {2016},
  doi     = {10.1145/2955101}
}

@inproceedings{sarwar2001,
  author    = {Sarwar, Badrul and Karypis, George and Konstan, Joseph and Riedl, John},
  title     = {Item-Based Collaborative Filtering Recommendation Algorithms},
  booktitle = {Proceedings of the 10th International Conference on World Wide Web},
  pages     = {285--295},
  year      = {2001},
  publisher = {ACM},
  doi       = {10.1145/371920.372071}
}

@inproceedings{klenitskiy2023,
  author    = {Klenitskiy, Anton and Vasilev, Alexey},
  title     = {Turning Dross Into Gold Loss: Is {BERT4Rec} Really Better than {SASRec}?},
  booktitle = {Proceedings of the 17th ACM Conference on Recommender Systems},
  pages     = {1120--1125},
  year      = {2023},
  publisher = {ACM},
  doi       = {10.1145/3604915.3610644}
}

@article{geirhos2020shortcut,
  author  = {Geirhos, Robert and Jacobsen, J{\"o}rn-Henrik and Michaelis, Claudio and Zemel, Richard and Brendel, Wieland and Bethge, Matthias and Wichmann, Felix A.},
  title   = {Shortcut Learning in Deep Neural Networks},
  journal = {Nature Machine Intelligence},
  volume  = {2},
  number  = {11},
  pages   = {665--673},
  year    = {2020},
  doi     = {10.1038/s42256-020-00257-z}
}

@misc{han2026shortcut,
  author        = {Han, Haoyu and Ma, Li and Wang, Hanbing and Li, Bingheng and Zha, Daochen and Tan, Chun How and Gao, Huiji and Liu, Xin and Moyerman, Stephanie and Katariya, Sanjeev and Liu, Hui and Tang, Jiliang},
  title         = {An Embarrassingly Simple Graph Heuristic Reveals Shortcut-Solvable Benchmarks for Sequential Recommendation},
  year          = {2026},
  eprint        = {2605.07125},
  archivePrefix = {arXiv},
  primaryClass  = {cs.IR}
}

@misc{cheng2026unier,
  author        = {Cheng, Xinghe and Zhuang, Guiyong and Xie, Yusheng and Wang, Jiapu and Liu, Yixin and Guan, Quanlong and Fang, Liangda and Pan, Shirui},
  title         = {{UniER}: A Unified Benchmark for Item-level and Path-level Exercise Recommendation},
  year          = {2026},
  eprint        = {2605.16750},
  archivePrefix = {arXiv},
  primaryClass  = {cs.IR}
}

@inproceedings{chaney2018,
  author    = {Chaney, Allison J. B. and Stewart, Brandon M. and Engelhardt, Barbara E.},
  title     = {How Algorithmic Confounding in Recommendation Systems Increases Homogeneity and Decreases Utility},
  booktitle = {Proceedings of the 12th ACM Conference on Recommender Systems},
  pages     = {224--232},
  year      = {2018},
  publisher = {ACM},
  doi       = {10.1145/3240323.3240370}
}

@inproceedings{jeunen2019,
  author    = {Jeunen, Olivier},
  title     = {Revisiting Offline Evaluation for Implicit-Feedback Recommender Systems},
  booktitle = {Proceedings of the 13th ACM Conference on Recommender Systems},
  pages     = {596--600},
  year      = {2019},
  publisher = {ACM},
  doi       = {10.1145/3298689.3347069}
}

@misc{schnabel2016,
  author        = {Schnabel, Tobias and Swaminathan, Adith and Singh, Ashudeep and Chandak, Navin and Joachims, Thorsten},
  title         = {Recommendations as Treatments: Debiasing Learning and Evaluation},
  year          = {2016},
  eprint        = {1602.05352},
  archivePrefix = {arXiv}
}

@article{erdt2015,
  author  = {Erdt, Mojisola and Fernandez, Alejandro and Rensing, Christoph},
  title   = {Evaluating Recommender Systems for Technology Enhanced Learning: A Quantitative Survey},
  journal = {IEEE Transactions on Learning Technologies},
  volume  = {8},
  number  = {4},
  pages   = {326--344},
  year    = {2015},
  doi     = {10.1109/TLT.2015.2438867}
}

@incollection{manouselis2011,
  author    = {Manouselis, Nikos and Drachsler, Hendrik and Vuorikari, Riina and Hummel, Hans and Koper, Rob},
  title     = {Recommender Systems in Technology Enhanced Learning},
  booktitle = {Recommender Systems Handbook},
  pages     = {387--415},
  year      = {2011},
  publisher = {Springer},
  doi       = {10.1007/978-0-387-85820-3_12}
}

@incollection{drachsler2015,
  author    = {Drachsler, Hendrik and Verbert, Katrien and Santos, Olga C. and Manouselis, Nikos},
  title     = {Panorama of Recommender Systems to Support Learning},
  booktitle = {Recommender Systems Handbook},
  pages     = {421--451},
  year      = {2015},
  publisher = {Springer},
  doi       = {10.1007/978-1-4899-7637-6_12}
}

@inproceedings{doroudi2017rem,
  author    = {Doroudi, Shayan and Aleven, Vincent and Brunskill, Emma},
  title     = {Robust Evaluation Matrix: Towards a More Principled Offline Exploration of Instructional Policies},
  booktitle = {Proceedings of the Fourth (2017) ACM Conference on Learning @ Scale},
  pages     = {3--12},
  year      = {2017},
  publisher = {ACM},
  doi       = {10.1145/3051457.3051463}
}

@article{doroudi2019reward,
  author  = {Doroudi, Shayan and Aleven, Vincent and Brunskill, Emma},
  title   = {Where's the Reward? {A} Review of Reinforcement Learning for Instructional Sequencing},
  journal = {International Journal of Artificial Intelligence in Education},
  volume  = {29},
  number  = {4},
  pages   = {568--620},
  year    = {2019},
  doi     = {10.1007/s40593-019-00187-x}
}

@inproceedings{pardos2017,
  author    = {Pardos, Zachary A. and Tang, Steven and Davis, Daniel and Le, Christopher Vu},
  title     = {Enabling Real-Time Adaptivity in {MOOCs} with a Personalized Next-Step Recommendation Framework},
  booktitle = {Proceedings of the Fourth (2017) ACM Conference on Learning @ Scale},
  pages     = {23--32},
  year      = {2017},
  publisher = {ACM},
  doi       = {10.1145/3051457.3051471}
}

@misc{tang2016mooc,
  author        = {Tang, Steven and Peterson, Joshua C. and Pardos, Zachary A.},
  title         = {Modelling Student Behavior using Granular Large Scale Action Data from a {MOOC}},
  year          = {2016},
  eprint        = {1608.04789},
  archivePrefix = {arXiv}
}

@misc{ma2025llmcourse,
  author        = {Ma, Boxuan and Khan, Md Akib Zabed and Yang, Tianyuan and Polyzou, Agoritsa and Konomi, Shin'ichi},
  title         = {How Good Are Large Language Models for Course Recommendation in {MOOCs}?},
  year          = {2025},
  eprint        = {2504.08208},
  archivePrefix = {arXiv},
  primaryClass  = {cs.IR}
}

@inproceedings{hou2024zeroshot,
  author    = {Hou, Yupeng and Zhang, Junjie and Lin, Zihan and Lu, Hongyu and Xie, Ruobing and McAuley, Julian and Zhao, Wayne Xin},
  title     = {Large Language Models are Zero-Shot Rankers for Recommender Systems},
  booktitle = {Advances in Information Retrieval (ECIR 2024)},
  series    = {Lecture Notes in Computer Science},
  pages     = {364--381},
  year      = {2024},
  publisher = {Springer},
  doi       = {10.1007/978-3-031-56060-6_24}
}

@misc{zheng2023mcq,
  author        = {Zheng, Chujie and Zhou, Hao and Meng, Fandong and Zhou, Jie and Huang, Minlie},
  title         = {Large Language Models Are Not Robust Multiple Choice Selectors},
  year          = {2023},
  eprint        = {2309.03882},
  archivePrefix = {arXiv}
}

@article{feng2009,
  author  = {Feng, Mingyu and Heffernan, Neil and Koedinger, Kenneth},
  title   = {Addressing the Assessment Challenge with an Online System that Tutors as it Assesses},
  journal = {User Modeling and User-Adapted Interaction},
  volume  = {19},
  number  = {3},
  pages   = {243--266},
  year    = {2009},
  doi     = {10.1007/s11257-009-9063-7}
}

@article{heffernan2014,
  author  = {Heffernan, Neil T. and Heffernan, Cristina Lindquist},
  title   = {The {ASSISTments} Ecosystem: Building a Platform that Brings Scientists and Teachers Together for Minimally Invasive Research on Human Learning and Teaching},
  journal = {International Journal of Artificial Intelligence in Education},
  volume  = {24},
  number  = {4},
  pages   = {470--497},
  year    = {2014},
  doi     = {10.1007/s40593-014-0024-x}
}

@inproceedings{choi2020ednet,
  author    = {Choi, Youngduck and Lee, Youngnam and Shin, Dongmin and Cho, Junghyun and Park, Seoyon and Lee, Seewoo and Baek, Jineon and Bae, Chan and Kim, Byungsoo and Heo, Jaewe},
  title     = {{EdNet}: A Large-Scale Hierarchical Dataset in Education},
  booktitle = {Artificial Intelligence in Education (AIED 2020)},
  series    = {Lecture Notes in Computer Science},
  pages     = {69--73},
  year      = {2020},
  publisher = {Springer},
  doi       = {10.1007/978-3-030-52240-7_13}
}

@article{corbett1994,
  author  = {Corbett, Albert T. and Anderson, John R.},
  title   = {Knowledge Tracing: Modeling the Acquisition of Procedural Knowledge},
  journal = {User Modeling and User-Adapted Interaction},
  volume  = {4},
  number  = {4},
  pages   = {253--278},
  year    = {1994},
  doi     = {10.1007/BF01099821}
}

@article{wilson2019eighty,
  author  = {Wilson, Robert C. and Shenhav, Amitai and Straccia, Mark and Cohen, Jonathan D.},
  title   = {The Eighty Five Percent Rule for Optimal Learning},
  journal = {Nature Communications},
  volume  = {10},
  number  = {1},
  pages   = {4646},
  year    = {2019},
  doi     = {10.1038/s41467-019-12552-4}
}

@misc{wang2020eedi,
  author        = {Wang, Zichao and Lamb, Angus and Saveliev, Evgeny and Cameron, Pashmina and Zaykov, Yordan and Hern{\'a}ndez-Lobato, Jos{\'e} Miguel and Turner, Richard E. and Baraniuk, Richard G. and Barton, Craig and Peyton Jones, Simon and Woodhead, Simon and Zhang, Cheng},
  title         = {Instructions and Guide for Diagnostic Questions: The {NeurIPS} 2020 Education Challenge},
  year          = {2020},
  eprint        = {2007.12061},
  archivePrefix = {arXiv},
  primaryClass  = {cs.CY}
}

@misc{qwen2024,
  author        = {{Qwen Team}},
  title         = {{Qwen2.5} Technical Report},
  year          = {2024},
  eprint        = {2412.15115},
  archivePrefix = {arXiv}
}

@misc{ozyurt2025exrec,
  author        = {Ozyurt, Yilmazcan and Almaci, Tunaberk and Feuerriegel, Stefan and Sachan, Mrinmaya},
  title         = {Personalized Exercise Recommendation with Semantically-Grounded Knowledge Tracing},
  year          = {2025},
  eprint        = {2507.11060},
  archivePrefix = {arXiv}
}

@misc{hooshyar2025llm,
  author        = {Hooshyar, Danial and Yang, Yeongwook and {\v{S}}{\'\i}r, Gustav and K{\"a}rkk{\"a}inen, Tommi and H{\"a}m{\"a}l{\"a}inen, Raija and Cukurova, Mutlu and Azevedo, Roger},
  title         = {Problems With Large Language Models for Learner Modelling: Why {LLMs} Alone Fall Short for Responsible Tutoring in {K}--12 Education},
  year          = {2025},
  eprint        = {2512.23036},
  archivePrefix = {arXiv}
}

@article{jdidou2016,
  author  = {Jdidou, Youssef and Khaldi, Mohamed},
  title   = {Increasing the Profitability of Students in {MOOCs} Using Recommendation Systems},
  journal = {International Journal of Knowledge Society Research},
  volume  = {7},
  number  = {4},
  pages   = {75--85},
  year    = {2016},
  doi     = {10.4018/IJKSR.2016100107}
}

@article{jdidou2021,
  author  = {Jdidou, Youssef and Aammou, Souhaib and Khaldi, Mohamed},
  title   = {Adapt Learning Path by Recommending Problems to Struggling Learners},
  journal = {International Journal of Emerging Technologies in Learning (iJET)},
  volume  = {16},
  number  = {20},
  pages   = {163--178},
  year    = {2021},
  doi     = {10.3991/ijet.v16i20.24283}
}

@article{gervet2020,
  author  = {Gervet, Theophile and Koedinger, Ken and Schneider, Jeff and Mitchell, Tom},
  title   = {When is Deep Learning the Best Approach to Knowledge Tracing?},
  journal = {Journal of Educational Data Mining},
  volume  = {12},
  number  = {3},
  pages   = {31--54},
  year    = {2020},
  doi     = {10.5281/zenodo.4143614}
}

@inproceedings{chang2015junyi,
  author    = {Chang, Haw-Shiuan and Hsu, Hwai-Jung and Chen, Kuan-Ta},
  title     = {Modeling Exercise Relationships in E-Learning: A Unified Approach},
  booktitle = {Proceedings of the 8th International Conference on Educational Data Mining (EDM 2015)},
  address   = {Madrid, Spain},
  year      = {2015}
}

@article{ludewig2018evaluation,
  author  = {Ludewig, Malte and Jannach, Dietmar},
  title   = {Evaluation of Session-Based Recommendation Algorithms},
  journal = {User Modeling and User-Adapted Interaction},
  volume  = {28},
  number  = {4-5},
  pages   = {331--390},
  year    = {2018},
  doi     = {10.1007/s11257-018-9209-6}
}

@inproceedings{jannach2017rnnknn,
  author    = {Jannach, Dietmar and Ludewig, Malte},
  title     = {When Recurrent Neural Networks Meet the Neighborhood for Session-Based Recommendation},
  booktitle = {Proceedings of the Eleventh ACM Conference on Recommender Systems},
  pages     = {306--310},
  year      = {2017},
  doi       = {10.1145/3109859.3109872}
}

@inproceedings{piech2015dkt,
  author    = {Piech, Chris and Bassen, Jonathan and Huang, Jonathan and Ganguli, Surya and Sahami, Mehran and Guibas, Leonidas J. and Sohl-Dickstein, Jascha},
  title     = {Deep Knowledge Tracing},
  booktitle = {Advances in Neural Information Processing Systems 28},
  pages     = {505--513},
  year      = {2015}
}

@inproceedings{kumar2019jodie,
  author    = {Kumar, Srijan and Zhang, Xikun and Leskovec, Jure},
  title     = {Predicting Dynamic Embedding Trajectory in Temporal Interaction Networks},
  booktitle = {Proceedings of the 25th ACM SIGKDD International Conference on Knowledge Discovery \& Data Mining},
  pages     = {1269--1278},
  year      = {2019},
  doi       = {10.1145/3292500.3330895}
}

@misc{jdidou2026code,
  author       = {Jdidou, Youssef},
  title        = {beat-the-syllabus: Structural Baselines and Lift-over-Structure Reports for Educational Recommender Systems},
  year         = {2026},
  howpublished = {Software, version 1.0.0, Zenodo},
  doi          = {10.5281/zenodo.24443552},
  url          = {https://github.com/jdidou/beat-the-syllabus}
}

@article{harris2020numpy,
  author  = {Harris, Charles R. and Millman, K. Jarrod and van der Walt, St{\'e}fan J. and others},
  title   = {Array Programming with {NumPy}},
  journal = {Nature},
  volume  = {585},
  number  = {7825},
  pages   = {357--362},
  year    = {2020},
  doi     = {10.1038/s41586-020-2649-2}
}

@article{virtanen2020scipy,
  author  = {Virtanen, Pauli and Gommers, Ralf and Oliphant, Travis E. and others},
  title   = {{SciPy} 1.0: Fundamental Algorithms for Scientific Computing in {Python}},
  journal = {Nature Methods},
  volume  = {17},
  number  = {3},
  pages   = {261--272},
  year    = {2020},
  doi     = {10.1038/s41592-019-0686-2}
}

@inproceedings{paszke2019pytorch,
  author    = {Paszke, Adam and Gross, Sam and Massa, Francisco and others},
  title     = {{PyTorch}: An Imperative Style, High-Performance Deep Learning Library},
  booktitle = {Advances in Neural Information Processing Systems 32},
  year      = {2019},
  eprint        = {1912.01703},
  archivePrefix = {arXiv}
}

@inproceedings{wolf2020transformers,
  author    = {Wolf, Thomas and Debut, Lysandre and Sanh, Victor and others},
  title     = {Transformers: State-of-the-Art Natural Language Processing},
  booktitle = {Proceedings of the 2020 Conference on Empirical Methods in Natural Language Processing: System Demonstrations},
  pages     = {38--45},
  publisher = {Association for Computational Linguistics},
  year      = {2020},
  doi       = {10.18653/v1/2020.emnlp-demos.6}
}

@misc{implicit,
  author       = {Frederickson, Ben},
  title        = {Implicit: Fast {Python} Collaborative Filtering for Implicit Datasets},
  howpublished = {Software, version 0.7.3},
  year         = {2026},
  note         = {Accessed November 2026},
  url          = {https://github.com/benfred/implicit}
}

@misc{llamacpp,
  author       = {{ggml-org}},
  title        = {llama.cpp: {LLM} Inference in {C/C++}},
  howpublished = {Software, used through llama-cpp-python 0.3.19},
  year         = {2026},
  note         = {Accessed November 2026},
  url          = {https://github.com/ggml-org/llama.cpp}
}

\appendix

\section{Hyperparameters and implementation details}\label{app:hparams}
All methods were implemented in Python (NumPy \citep{harris2020numpy}, SciPy \citep{virtanen2020scipy}, PyTorch \citep{paszke2019pytorch}; iALS with the \texttt{implicit} library \citep{implicit}) and run on CPUs with two threads per run (four for the 3B and 7B LLMs). For each personalised method, the configuration with the highest validation NDCG@10 was selected (Table~\ref{tab:grids}); the structural predictors have no hyperparameters.

\begin{table}[h]
\centering
\small
\caption{Hyperparameter grids. History vectors use the learner's 50 most recent distinct activities (200 for EASE); item--item statistics use each training learner's 100 most recent distinct activities (200 for EASE).}
\label{tab:grids}
\resizebox{\linewidth}{!}{%
\begin{tabular}{ll}
\toprule
Method & Grid and fixed settings \\
\midrule
ItemKNN & neighbours $k\in\{50,200\}$; recency decay $\gamma\in\{1.0,0.7\}$; cosine similarity with shrinkage 10 \\
RP3beta & $\alpha\in\{0.5,1.0\}$; $\beta\in\{0,0.3,0.6\}$; 200 neighbours per activity \\
EASE & $\lambda\in\{50,200,800\}$; restricted to the 8,000 most popular activities, the others ranked after them \\
iALS & 64 factors; regularisation $\in\{0.01,0.1\}$; confidence $\alpha\in\{1,10\}$; 15 iterations; history folded in \\
MCLD & original rule: all most-coincident peers, least-different tier, then further tiers and popularity \\
aMCLD & number of peers $k\in\{1,5,20,50,200\}$ in the order (coincidence $\downarrow$, difference $\uparrow$) \\
SKNN & neighbours $k\in\{50,200\}$; session of the $m\in\{5,20\}$ most recent distinct activities, linearly decaying weights; cosine similarity \\
ZPD & target success rate $\tau\in\{0.5,0.6,0.7,0.8,0.9\}$ \\
DKT & $\tau$ as ZPD; LSTM with 64 units over (skill, outcome) steps, windows of 200 steps, Adam ($10^{-3}$), batch 64, \\
 & at most 20 epochs or 5 minutes per training split; activities without a skill share one extra skill \\
SASRec & $d=64$, 2 layers, 2 heads, length 50, dropout 0.2, Adam (learning rate $10^{-3}$), batch 256; \\
 & at most 200 epochs, early stopping on validation NDCG@10 (every 2 epochs, patience 4); \\
 & full softmax up to 3,000 activities, otherwise sampled softmax with 2,048 uniform negatives \\
\bottomrule
\end{tabular}}
\end{table}

\paragraph{MCLD on large cohorts.} The original MCLD selects peers among the classmates of a course run. Here every training learner is a potential peer, the learner's own training sequence excluded. Remediation, which re-proposes failed activities first, plays no role because the target is always a new activity.

\paragraph{Computational shortcuts.} Item--item statistics are computed in blocks and pruned to the top neighbours of each activity. EASE is restricted to the 8,000 most popular activities on the three largest catalogues (A09, A12 and EdNet), which cover most of their steps. Each SASRec fit had a budget of 75 minutes under protocol P1 (30 minutes under P2, a robustness check); early stopping on a random subset of 2,000 validation cases decided the number of epochs, and the selected model was then evaluated on all validation cases. Under P2, the model was then retrained on the whole training period for the selected number of epochs. The P1 budget was reached on \numsasCapList{}.

\section{Selection on validation}\label{app:valsel}
The best structural predictor and the best personalised recommender of Tables~\ref{tab:lift} and~\ref{tab:p2} are chosen by their test scores, for both sides alike. Table~\ref{tab:valsel} chooses them by their validation scores instead. The choice changes on \numvalChanged{} dataset; the personalised recommender is significantly better on \numvalSigList{} and significantly worse on \numvalWorseList{}, as in Table~\ref{tab:lift}, and the structural share ranges from \numvalRhoMin{} to \numvalRhoMax{} (median \numvalRhoMedian{}).
\begin{table}[h]
\centering
\caption{Lift over structure under protocol P1 when the best structural predictor and the best personalised recommender are chosen by their validation score instead of their test score (NDCG@10). Notation as in Table~\ref{tab:lift}; $\rho$ is here the best structural score divided by the larger of the two scores.}
\label{tab:valsel}
\small
\resizebox{\linewidth}{!}{\begin{tabular}{llrlrrrr}
\toprule
Dataset & Best structural & $S_{\text{struct}}$ & Best personalised & Score & $\Delta$ [95\% interval] & $p_{\text{Holm}}$ & $\rho$ \\
\midrule
A09 & CONT & 0.083 & SASRec & 0.077 & -0.006 [-0.013, +0.002] & 0.304 & 1.00 \\
A12 & MC & 0.276 & SASRec & 0.256 & -0.020 [-0.023, -0.017] & $<$0.001 & 1.00 \\
A15 & MC & 0.420 & SASRec & 0.522 & +0.102 [+0.096, +0.108] & $<$0.001 & 0.81 \\
A17 & MC & 0.713 & SASRec & 0.728 & +0.015 [+0.008, +0.022] & $<$0.001 & 0.98 \\
EdNet & MC & 0.035 & SASRec & 0.042 & +0.007 [+0.003, +0.012] & $<$0.001 & 0.82 \\
Junyi & MC & 0.488 & SASRec & 0.535 & +0.047 [+0.044, +0.051] & $<$0.001 & 0.91 \\
Eedi & MC & 0.558 & SASRec & 0.559 & +0.001 [-0.006, +0.008] & 0.779 & 1.00 \\
MOOC & MC & 0.692 & SASRec & 0.780 & +0.088 [+0.083, +0.093] & $<$0.001 & 0.89 \\
\bottomrule
\end{tabular}
}
\end{table}

\section{More results under protocol P1}\label{app:more}
\figmain
\begin{table}[t]
\centering
\caption{On-path and off-path test steps (protocol P1, NDCG@10). On-path: the next activity is in the container of the last one; off-path: it is not. \emph{Struct.} and \emph{Pers.}: best structural predictor and best personalised recommender within each group. $\Delta$ off-path: lift of the best personalised over the best structural predictor on off-path steps.}
\label{tab:paths}
\small
\setlength{\tabcolsep}{4pt}
\resizebox{\linewidth}{!}{\begin{tabular}{lrrrlrrlr}
\toprule
 & & \multicolumn{3}{c}{On-path steps} & \multicolumn{4}{c}{Off-path steps} \\
\cmidrule(lr){3-5}\cmidrule(lr){6-9}
Dataset & Off-path & Struct. & Pers. & Best pers. & Struct. & Pers. & Best pers. & $\Delta$ off-path \\
\midrule
A09 & 37\% & 0.132 & 0.107 & ItemKNN & 0.033 & 0.030 & SASRec & -0.004 \\
A12 & 31\% & 0.257 & 0.239 & SASRec & 0.319 & 0.294 & SASRec & -0.024 \\
A17 & 60\% & 0.968 & 0.969 & SASRec & 0.544 & 0.568 & SASRec & +0.024 \\
EdNet & 23\% & 0.034 & 0.044 & SASRec & 0.037 & 0.049 & SKNN & +0.011 \\
Junyi & 40\% & 0.748 & 0.792 & SASRec & 0.130 & 0.169 & EASE & +0.039 \\
Eedi & 7\% & 0.581 & 0.579 & SASRec & 0.254 & 0.298 & SASRec & +0.045 \\
\bottomrule
\end{tabular}
}
\end{table}

\section{Variation between training runs}\label{app:seeds}
SASRec was trained three more times with other seeds (on the author's computer, two threads per run) on the \numseedN{} datasets where its lift matters most. Table~\ref{tab:seeds} gives its NDCG@10 in the main run and in each further run, and its lift over the best structural predictor on the same test cases, with the number of runs whose paired bootstrap 95\% interval excludes zero.
\begin{table}[h]
\centering
\caption{SASRec over four training runs (protocol P1, NDCG@10): the main run of Table~\ref{tab:main} and three runs with seeds 1 to 3. $\Delta$ range: lift over the best structural predictor across the four runs; Significant: runs whose paired bootstrap 95\% interval excludes zero.}
\label{tab:seeds}
\small
\resizebox{\linewidth}{!}{\begin{tabular}{llrrrrrrrl}
\toprule
Dataset & Best structural & Score & Main run & Seed 1 & Seed 2 & Seed 3 & Mean (sd) & $\Delta$ range & Significant \\
\midrule
A15 & MC & 0.420 & 0.522 & 0.520 & 0.531 & 0.529 & 0.525 (0.005) & +0.099 to +0.110 & 4 of 4 \\
A17 & MC & 0.713 & 0.728 & 0.722 & 0.732 & 0.723 & 0.726 (0.005) & +0.009 to +0.019 & 4 of 4 \\
Junyi & MC & 0.488 & 0.535 & 0.536 & 0.534 & 0.534 & 0.535 (0.001) & +0.046 to +0.048 & 4 of 4 \\
Eedi & MC & 0.558 & 0.559 & 0.592 & 0.601 & 0.582 & 0.584 (0.018) & +0.001 to +0.043 & 3 of 4 \\
MOOC & MC & 0.692 & 0.780 & 0.776 & 0.787 & 0.775 & 0.780 (0.005) & +0.083 to +0.094 & 4 of 4 \\
\bottomrule
\end{tabular}
}
\end{table}

\section{Full results under protocol P2}\label{app:p2}
\begin{table}[t]
\centering
\caption{Lift over structure under protocol P2 (global temporal split), NDCG@10. Notation as in Table~\ref{tab:lift}; the full results are in Appendix~\ref{app:p2}.}
\label{tab:p2}
\small
\resizebox{\linewidth}{!}{\begin{tabular}{llrlrrrrr}
\toprule
Dataset & Best structural & $S_{\text{struct}}$ & Best personalised & Score & $\Delta$ [95\% interval] & $p_{\text{Holm}}$ & $\rho$ & $\rho_{\text{meta}}$ \\
\midrule
A09 & CONT & 0.024 & ItemKNN & 0.010 & -0.015 [-0.036, +0.002] & 0.113 & 1.00 & 1.00 \\
A12 & MC & 0.033 & SASRec & 0.027 & -0.006 [-0.008, -0.004] & $<$0.001 & 1.00 & 0.15 \\
A15 & MC & 0.232 & MCLD & 0.374 & +0.142 [+0.127, +0.156] & $<$0.001 & 0.62 & 0.26 \\
A17 & MC & 0.309 & SASRec & 0.322 & +0.012 [+0.004, +0.021] & 0.006 & 0.96 & 0.38 \\
EdNet & MC & 0.013 & ItemKNN & 0.020 & +0.007 [+0.000, +0.014] & 0.074 & 0.65 & 0.33 \\
Junyi & MC & 0.149 & SASRec & 0.176 & +0.027 [+0.018, +0.036] & $<$0.001 & 0.85 & 0.48 \\
Eedi & MC & 0.382 & SASRec & 0.411 & +0.029 [+0.021, +0.036] & $<$0.001 & 0.93 & 0.56 \\
MOOC & MC & 0.610 & SASRec & 0.682 & +0.071 [+0.063, +0.080] & $<$0.001 & 0.90 & 0.41 \\
\bottomrule
\end{tabular}
}
\end{table}
\begin{table}[h]
\centering
\caption{NDCG@10 for the next new activity under protocol P2 (global temporal split). Bold: best method; underlined: best structural predictor.}
\label{tab:p2full}
\small
\resizebox{\linewidth}{!}{\begin{tabular}{lrrrrrrrr}
\toprule
Method & A09 & A12 & A15 & A17 & EdNet & Junyi & Eedi & MOOC \\
\midrule
\multicolumn{9}{l}{\emph{Reference}} \\
RAND & 0.000 & 0.000 & 0.049 & 0.002 & 0.001 & 0.007 & 0.065 & 0.073 \\
\multicolumn{9}{l}{\emph{Structural (not personalised)}} \\
POP & 0.000 & 0.001 & 0.097 & 0.005 & 0.007 & 0.076 & 0.186 & 0.229 \\
ORDER & 0.000 & 0.001 & 0.069 & 0.014 & 0.001 & 0.027 & 0.210 & 0.278 \\
CONT & \underline{\textbf{0.024}} & 0.005 & 0.069 & 0.124 & 0.005 & 0.084 & 0.229 & 0.278 \\
MC & 0.023 & \underline{\textbf{0.033}} & \underline{0.232} & \underline{0.309} & \underline{0.013} & \underline{0.149} & \underline{0.382} & \underline{0.610} \\
\multicolumn{9}{l}{\emph{Personalised}} \\
ItemKNN & 0.010 & 0.025 & 0.218 & 0.274 & \textbf{0.020} & 0.121 & 0.277 & 0.435 \\
RP3beta & 0.008 & 0.022 & 0.221 & 0.100 & 0.017 & 0.109 & 0.190 & 0.277 \\
EASE & 0.002 & 0.023 & 0.224 & 0.176 & 0.016 & 0.130 & 0.237 & 0.363 \\
iALS & 0.000 & 0.016 & 0.154 & 0.058 & 0.011 & 0.108 & 0.263 & 0.375 \\
SKNN & 0.009 & 0.022 & 0.305 & 0.274 & 0.017 & 0.163 & 0.352 & 0.403 \\
MCLD & 0.000 & 0.005 & \textbf{0.374} & 0.009 & 0.001 & 0.092 & 0.235 & 0.272 \\
aMCLD & 0.000 & 0.015 & 0.318 & 0.063 & 0.006 & 0.097 & 0.229 & 0.302 \\
ZPD & 0.000 & 0.000 & 0.044 & 0.002 & 0.000 & 0.030 & 0.048 & -- \\
DKT & 0.000 & 0.000 & 0.058 & 0.002 & 0.001 & 0.023 & 0.059 & -- \\
SASRec & 0.007 & 0.027 & 0.335 & \textbf{0.322} & 0.017 & \textbf{0.176} & \textbf{0.411} & \textbf{0.682} \\
\bottomrule
\end{tabular}
}
\end{table}

\section{HR@10 under protocol P1}\label{app:hr10}
\begin{table}[h]
\centering
\caption{HR@10 for the next new activity under protocol P1. Bold: best method; underlined: best structural predictor.}
\label{tab:hr10}
\small
\resizebox{\linewidth}{!}{\begin{tabular}{lrrrrrrrr}
\toprule
Method & A09 & A12 & A15 & A17 & EdNet & Junyi & Eedi & MOOC \\
\midrule
\multicolumn{9}{l}{\emph{Reference}} \\
RAND & 0.001 & 0.000 & 0.109 & 0.004 & 0.001 & 0.015 & 0.144 & 0.157 \\
\multicolumn{9}{l}{\emph{Structural (not personalised)}} \\
POP & 0.009 & 0.005 & 0.213 & 0.018 & 0.016 & 0.234 & 0.414 & 0.514 \\
ORDER & 0.004 & 0.010 & 0.133 & 0.098 & 0.002 & 0.131 & 0.503 & 0.702 \\
CONT & \underline{\textbf{0.173}} & 0.125 & 0.133 & 0.341 & 0.021 & 0.438 & 0.593 & 0.702 \\
MC & 0.147 & \underline{\textbf{0.346}} & \underline{0.631} & \underline{0.779} & \underline{0.067} & \underline{0.631} & \underline{0.824} & \underline{0.957} \\
\multicolumn{9}{l}{\emph{Personalised}} \\
ItemKNN & 0.146 & 0.298 & 0.626 & 0.730 & 0.075 & 0.555 & 0.715 & 0.827 \\
RP3beta & 0.095 & 0.224 & 0.638 & 0.299 & 0.047 & 0.560 & 0.560 & 0.674 \\
EASE & 0.088 & 0.135 & 0.679 & 0.550 & 0.044 & 0.618 & 0.722 & 0.881 \\
iALS & 0.058 & 0.101 & 0.481 & 0.130 & 0.034 & 0.549 & 0.673 & 0.770 \\
SKNN & 0.123 & 0.273 & 0.702 & 0.625 & 0.071 & 0.597 & 0.793 & 0.868 \\
MCLD & 0.052 & 0.066 & 0.641 & 0.012 & 0.002 & 0.395 & 0.533 & 0.717 \\
aMCLD & 0.047 & 0.126 & 0.634 & 0.052 & 0.018 & 0.436 & 0.517 & 0.711 \\
ZPD & 0.007 & 0.002 & 0.130 & 0.004 & 0.002 & 0.173 & 0.190 & -- \\
DKT & 0.002 & 0.000 & 0.130 & 0.017 & 0.001 & 0.033 & 0.129 & -- \\
SASRec & 0.139 & 0.315 & \textbf{0.730} & \textbf{0.786} & \textbf{0.083} & \textbf{0.671} & \textbf{0.827} & \textbf{0.981} \\
\bottomrule
\end{tabular}
}
\end{table}

\section{LLM prompt}\label{app:prompt}
The text below is the prompt of one Junyi test case, verbatim. Each of the 20 candidate names is appended in turn as the continuation, and its mean token log-probability is computed; the same names are also scored after the first line and the last line alone (no history), and the difference gives the candidate's score. Activity names are those of the dataset, including its spelling.

\begin{Verbatim}
A student is learning on Junyi Academy, an online mathematics learning platform organised as a knowledge map. The activities are exercises.
The student's most recent activities, oldest first:
1. exercise 'subtraction 4' (topic: addition subtraction): passed
2. exercise 'addition 4' (topic: addition subtraction): passed
3. exercise 'subtraction 4' (topic: addition subtraction): not passed
4. exercise 'angle types' (topic: segments and angles): passed
5. exercise 'sides and angles of simple shapes' (topic: basic geometry): passed
6. exercise 'measuring and comparing angles' (topic: segments and angles): passed
7. exercise 'division 0.5' (topic: multiplication division): passed
8. exercise 'division 3' (topic: multiplication division): passed
9. exercise 'division 1.4' (topic: multiplication division): passed
10. exercise 'division 2.9' (topic: multiplication division): passed
The next activity the student starts is:

[candidate continuations, e.g.]
 exercise 'l ml transform' (topic: unit conversion)
 exercise 'recognizing shapes' (topic: basic geometry)
 exercise 'line relationships' (topic: linear equations and inequalitie)
 exercise 'number within fifty' (topic: addition subtraction)
 exercise 'adding decimals 2' (topic: decimals)
...
\end{Verbatim}

\end{document}